\documentclass{emulateapj}
\usepackage{apjfonts}
\usepackage{longtable}

\def\lya{Ly$\alpha$}
\def\lyb{Ly$\beta$}
\def\ly{Ly$\alpha$}
\def\nfv{N~{\sc v}}
\def\sitw{Si~{\sc ii}}
\def\onsitw{O~{\sc i}/Si~{\sc ii}}
\def\osx{O~{\sc vi}}
\def\ctw{C~{\sc ii}}
\def\sifrofr{Si~{\sc iv}/O~{\sc iv}]}
\def\cfr{C~{\sc iv}}
\def\cthr{C~{\sc iii}]}
\def\mgtw{Mg~{\sc ii}}
\def\nefva{[Ne~{\sc v}]$_1$}
\def\nefvb{[Ne~{\sc v}]$_2$}
\def\sithrfethr{Si~{\sc iii}]/Fe~{\sc iii}}
\def\althr{Al~{\sc iii}}

\slugcomment{}

\shorttitle{QSO Pair Spectroscopy}
\shortauthors{Marble et al.}
\journalinfo{Published in the Astrophysical Journal Supplement Series (2008, \apjs, 175, 29)}

\begin{document}

\title{The Flux Auto- and Cross-Correlation of the Ly$\alpha$ Forest.
\\I. Spectroscopy of QSO Pairs with Arcminute Separations and Similar Redshifts$^1$}
\footnotetext[1]{Observations reported here were obtained at the MMT Observatory,
    a joint facility of the University of Arizona and the Smithsonian
    Institution, and with the 6.5m Magellan Telescopes
    located at Las Campanas Observatory, Chile.}

\author{Andrew R. Marble, Kristoffer A. Eriksen, Chris D. Impey, Lei Bai}
\affil{Steward Observatory, University of Arizona, Tucson, AZ 85721}
\and
\author{Lance Miller}
\affil{Department of Physics, Oxford University, Keble Road, Oxford OX1 3RH, UK}

\begin{abstract}

The \lya\ forest has opened a new redshift regime for cosmological investigation.
At $z>2$ it provides a unique probe of cosmic geometry and an
independent constraint on dark energy that is not subject to standard
candle or ruler assumptions.
In Paper I of this series on using the \lya\ forest observed in pairs of QSOs 
for a new application of the Alcock-Paczy\'{n}ski test, 
we present and discuss the results of a campaign to
obtain moderate-resolution spectroscopy (FWHM $\simeq$ 2.5\AA) of the
\lya\ forest in pairs of QSOs with small redshift
differences ($\Delta z<0.25$, $z>2.2$) and arcminute separations 
($\theta<5\arcmin$).  
This data set,
composed of seven individual QSOs, 35 pairs, and one triplet,
is also well-suited for future investigations of the coherence of \lya\
absorbers on $\sim1$ Mpc transverse scales and the transverse
proximity effect.  We note seven revisions for previously published
QSO identifications and/or redshifts.
\end{abstract}


\section{Introduction}\label{intro}

Forty years after its existence was first predicted \citep{1965gun142apj1633,
1965sch207nat963, 1965shk8sva638,1965bah142apj1677}, the \lya\ forest has
emerged as a powerful cosmological tool
(see \citealt{1998rau36araa267} for a review).  Observed indirectly in 
the spectra of bright background sources, generally QSOs, non-uniform
neutral hydrogen gas along the line of sight results in varying
degrees of absorption caused by the redshifted \lya\ transition line.
The opacity of the gas maps onto the underlying mass distribution via
the fluctuating Gunn-Peterson model.  Remarkable agreement between cosmological
simulations and high-resolution observations has confirmed the
robustness of this relatively simple formalism
\citep{1994cen437apjl9, 1995zha453apjl57, 1996her457apjl51,
  1998the301mnras478}.

\lya\ forest observations along individual lines of sight are limited
in that they provide only one-dimensional information.  This can be 
extended to the transverse direction when two or more sufficiently
bright QSOs happen to lie in close angular proximity.
Approximately a dozen pairs and a triplet have been previously used 
to constrain the size and nature of \lya\ absorbers
\citep{1994bec437apjl83, 1994din437apjl87, 1996fan462apj77,
1998cro502apj16, 1998dod339aap678, 2000wil532apj77,
2000lis311mnras657, 2003rol341mnras1279}.
The 2dF QSO Redshift Survey \citep[2QZ;][]{2004cro349mnras1397} and Sloan
Digital Sky Survey \citep[SDSS;][]{2006ade162apjs38} have greatly
increased the number of known QSOs.  The fainter spectroscopic limiting
magnitude of 2QZ ($B<21$) has also resulted in the identification of relatively
rare pairs of QSOs with arcminute separations and similar redshifts, whereas follow-up
spectroscopy is generally required to confirm candidates
identified photometrically in SDSS.  In addition to studies of the
nature of \lya\ forest absorbers, such pairs are useful for the
transverse proximity effect and the Alcock-Paczy\'{n}ski (AP) test
for dark energy.

Motivated by the latter application, we have used the MMT and Magellan
telescopes to obtain science-grade spectra for QSO pairs with
arcminute separations and similar redshifts.  In \S\ref{pairs}
we define the sample criteria, and detail the observations and data
reduction in \S\S\ \ref{obs} and \ref{redux}, respectively.
The final data set is presented and discussed in \S\ref{sample}.
In \S\ref{errata} we
note seven revisions to identifications and/or redshifts
found in previously published QSO catalogues, before summarizing in
\S\ref{summary}.  
However, first we briefly discuss the
primary science motivators for this new data set.

\subsection{Alcock-Paczy\'{n}ski Test}\label{aptest}

If some supernovae are indeed true standard candles, then
observations of their light curves as a function of redshift reveal
that the expansion of the Universe is presently accelerating,
following a period of deceleration in the distant past
\citep{2003wan590apj944, 2003ton594apj1}.  This
inflection in the expansion rate provides the only direct evidence for
a nonzero cosmological constant, as $\Omega_{\Lambda}$ cannot be
measured directly from the cosmic microwave background
\citep{2003spe148apjs175}.  However, alternative models, characterized by supernova
with evolving properties and no cosmological constant, are also in
agreement with the supernovae data.  The merits of an independent
test for dark energy in a different redshift
regime ($z>2$) and subject to unrelated systematic errors are clear.

\citet{1999mcd518apj24} and \citet{1999hui511apjl5} proposed a new application
of the AP test \citep{1979alc281nat358}
using the \lya\ forests of QSO pairs separated by several arcminutes or
less. The AP test is a purely geometric method for measuring
cosmological parameters, which at $z>1$ is particularly sensitive to
$\Omega_{\Lambda}$. Simply stated, spherical objects
observed at high redshift will only appear spherical if the correct
cosmology is used to convert from angular to physical scales.
More generally, this test can be applied to the correlation function
of any isotropic tracer, such as the \lya\ forest.  
\citet{2003mcd585apj34} investigated this application of the AP test in
detail and found that neither high signal-to-noise ratio (S/N $>$ 10) nor high
resolution (FWHM $\lesssim$ 2\AA) are necessary.  Rather, a large number
of moderate quality data pairs is needed to reduce the effects of
sample variance (approximately $13\theta^2$ pairs with
separation less than $\theta$ arcminutes for a measurement of
$\Omega_\Lambda$ to within 5\%).
The more similar the redshifts of the paired QSOs, the
greater the overlap in the portion of the \lya\ forest unaffected by
\lyb\ absorption ($\Delta z \gtrsim 0.6$ provides no overlap at
$z=2.25$).

\subsection{Transverse Proximity Effect}\label{tpe}

There are two methods for measuring the intensity of the background
radiation responsible for ionizing the intergalactic medium. Provided
that the mean baryon density, gas temperature and power spectrum amplitude
are known, it can be derived from a comparison between numerical
simulations and the observed distribution of transmitted flux in the
\lya\ forest \citep{2001mcd549apj11, 2003sch584apj110}.  This approach
yields lesser values in disagreement with the proximity effect,
which measures the relative decrement in \lya\ absorption blueward of a
QSO's emission redshift caused by the increased ionizing background
from the QSO itself \citep{1982car198mnras91, 1986mur309apj19,
1988bat327apj570, 2000sco130apjs67, 2002scp571apj665}.

The transverse proximity effect is a variant on the latter method, in
which the decrement caused by a QSO is not measured from its own
spectrum but rather from the spectrum of a background QSO with a small
projected separation.  This has the advantage, assuming a sufficient
redshift difference [$\Delta z \gtrsim (1+z) / 30$], of moving the 
measurement away from the wings of
the broad \lya\ emission line where the continuum is often poorly
constrained.  To date, there have been several observational
studies involving 
three or fewer pairs \citep{1989cro336apj550, 1991dob377apjl69,
1992mol258aap234, 1995fer277mnras235, 2001lis328mnras653,
2003jak397aap891, 2004sch610apj105}, and no
unambiguous detection of the transverse proximity effect has been made.
This may be in part due to increased gas density associated with QSO
environments, as well as beaming and variability of QSO emission
\citep{2004sch610apj105}.  The effects of anisotropy and variability can
be investigated with a large sample of pairs spanning a range of
angular separations.  Ideally, these paired QSOs would be physically
unassociated ($\Delta v \gtrsim 2500$ km s$^{-1}$) but with
sufficiently small redshift differences ($\Delta z \lesssim 0.4$ for $z=2.25$)
that the region of analysis falls redward of the onset of \lyb\
absorption.

\section{Pair Selection}\label{pairs}

Our initial sample of known QSO pairs was drawn from the literature,
according to the following criteria tailored for the AP test. 
The angular separation was limited
to be less than five arcminutes (5.7 $h_{100}^{-1}$ comoving Mpc at $z=2.2$,
assuming $\Omega_m = 0.27$ and $\Omega_{\Lambda} = 0.73$) due to
the rapid decline in correlation of the \lya\ forest on
megaparsec scales.  In order to ensure overlapping coverage 
of the forest redward of the onset of Lyman$-\beta$ absorption, a
redshift difference of $\Delta z<0.25$ was required.  A minimum
redshift of $z>2.2$ was necessitated by diminishing
atmospheric transmission and spectrograph efficiency blueward of
3300\AA.  

We applied these criteria in 2002 to the 10$^{\mathrm{th}}$ edition of
the \citet{2001ver374aap92} catalogue of quasars and active
galactic nuclei (VCV) and the completed, but unpublished, 2QZ catalogue.
The limiting magnitude of the
latter ($B<21$) is well matched for our purposes: faint enough for
finding relatively rare QSO pairs, but bright enough for obtaining
spectra of sufficient quality with 6.5m telescopes.  The Sloan
Digital Sky Survey's spectroscopic QSO sample, on the other hand, has a
limiting magnitude of $i^{\prime}_{PSF}<19.1$.  A repeat query applied
to the 11$^{\mathrm{th}}$ edition of the VCV \citep{2003ver412aap399}
in late 2003 (which includes the full 2QZ release) resulted in
approximately 200 QSO pairs.

\section{Observations}\label{obs}

Optical spectroscopy was obtained in the northern and southern
hemispheres with the 6.5m MMT at Mount Hopkins, Arizona, and the
twin 6.5m Magellan telescopes at Las Campanas, Chile. All
observations were made between 2002 April 12 and 2004 May 18. 
Telescope scheduling and observing conditions largely determined which
QSOs were observed. However, when possible, precedence was
given to pairs with smaller separations and redshift differences.  In
the case of the triplet KP~1623+26.8A, KP~1623+26.8B, and
KP~1623.9+26.8, high quality spectra with superior resolution were
already available in the literature \citep{1989cro336apj550}; however, we
reobserved them in the interest of a homogeneous data set.

\subsection{MMT Configuration}\label{mmt}

All MMT data were taken with the Blue Channel spectrograph, the 800
grating, a 1$\arcsec\times$180$\arcsec$ slit and no filters. This
configuration yields a dispersion of 
0.75\AA\ per pixel, a nominal FWHM resolution of 2.2\AA\ and approximately
2000\AA\ of wavelength coverage redward of 3300\AA. 
For calibration purposes, the helium-argon
and neon arc lamps were used as well as the ``bright'' incandescent
lamp, all of which are located in the MMT Top Box.

\subsection{Magellan Configuration}\label{mag}

Observations during two runs at Magellan in 2002 and 2003 were made
with the Clay and 
Baade telescopes, respectively.  In both cases, the Boller \&
Chivens (B\&C) spectrograph was used, along with the 1200 grating blazed at
4000\AA, a 1$\arcsec\times$72$\arcsec$ slit and no filters.  The
resulting spectra span approximately 1600\AA\ redward of 3300\AA\ with 0.8\AA\ pixels and
have a nominal FWHM resolution of 2.4\AA.  In 2002, calibration was
carried out using the helium arc, argon arc and incandescent lamps
that are mounted to the exterior of the telescope and illuminate a
screen placed in front of the secondary mirror.  For the subsequent
run, helium and argon arc lamps located within the spectrograph
were used instead, due to their increased abundance and strength of
bluer lines.

\subsection{Modus Operandi}

The same protocol was followed for each night of observations.  Bias
frames, sky flats and flat field lamp images were taken during the lighter
hours prior to or following $12\degr$ twilight. Sky
flats were primarily necessary due to the significant difference in the chip
illumination patterns of the sky and the MMT Top Box incandescent
lamp. Multiple standard stars, characterized by minimal absorption and
spectral energy distributions peaking in the blue, were observed
between $12\degr$ and $18\degr$ twilight.

The QSOs were generally placed at the center of the slit; however, in
rare cases, both pair members were observed simultaneously.  
Exposure times were selected to yield a desired minimum
S/N of 10 per pixel in the final combined spectra.  Consecutive
exposures were made in order to identify outlier pixels affected by 
cosmic rays. Arc lamps were observed at each telescope pointing for
the most reliable wavelength calibration.  

Observing conditions were not generally photometric.  In many cases,
absolute flux calibration was affected by clouds and/or slit loss.
Relative flux calibration was benefited by observing QSOs at an
airmass less than 1.5 and with the slit at or near the parallactic
angle.  However, this minimizes, rather than eliminates,
wavelength-dependent light loss due to differential refraction and the
latter precaution is not applicable to cases in which the slit was
rotated to observe two QSOs simultaneously.  Because correlation
measurements in the \lya\ forest require the flux spectrum to
be normalized by the underlying QSO continuum, neither absolute nor
relative flux calibration errors affect the primary science objectives
of this data set.

\section{Data Reduction}\label{redux}

Data from all observing runs were processed in a uniform manner using
{\tt iSPEC2d}, a long-slit spectroscopy data reduction package
written in IDL. {\tt iSPEC2d} utilizes many standard techniques
similar to those found in other packages (\emph{e.g.}, IRAF),
as well as additional features including error propagation, bad pixel
tracking, two-dimensional sky subtraction and minimal interpolation.
Here we detail the data reduction steps followed for each night of
observing.

A bad pixel map was created in order to mask and exclude deviant
pixels/regions on the CCD.  A master bias frame was constructed from
$\gtrsim20$ 
zero-integration exposures by taking the median for each pixel after
discarding the highest and lowest two values.  A master sky flat and
master dome flat were constructed in the same manner after
normalizing the median counts of the corresponding exposures.
For the latter, approximately 100 exposures were used to
compensate for relatively few UV photons emitted by the incandescent
lamps.

Raw data frames were bias-subtracted, flat-fielded and illumination
corrected using the master calibration files.  Sky apertures were
interactively selected and two-dimensional sky subtraction was
performed, the advantages of which are discussed by
\citet{2003kel115pasp688}.  Rectification and wavelength calibration was
carried out using the two-dimensional wavelength solution
corresponding to the comparison lamp image taken at the same telescope 
pointing.  Atmospheric extinction and reddening as a function of
air mass were corrected for using either the Kitt Peak or CTIO extinction
curve.  Effective exposure times were calculated for observations made
through variable cloud cover, assuming wavelength-independent
obscuration.  This was done by normalizing the flux
rate of consecutive exposures to the highest such value.  Subexposures
were then combined, and, in the process, cosmic ray events were
identified and excluded.

A sensitivity curve, derived from standard stars observed at the
beginning or end of the night, was used for flux calibration.
One-dimensional spectra were then optimally extracted \citep{1986hor98pasp609}
using variance
weighting.  When relevant, multiple observations for the same target
were weighted by their effective exposure times and combined.

\section{Spectroscopic Sample}\label{sample}

\begin{figure}[t] 
\includegraphics[scale=0.96]{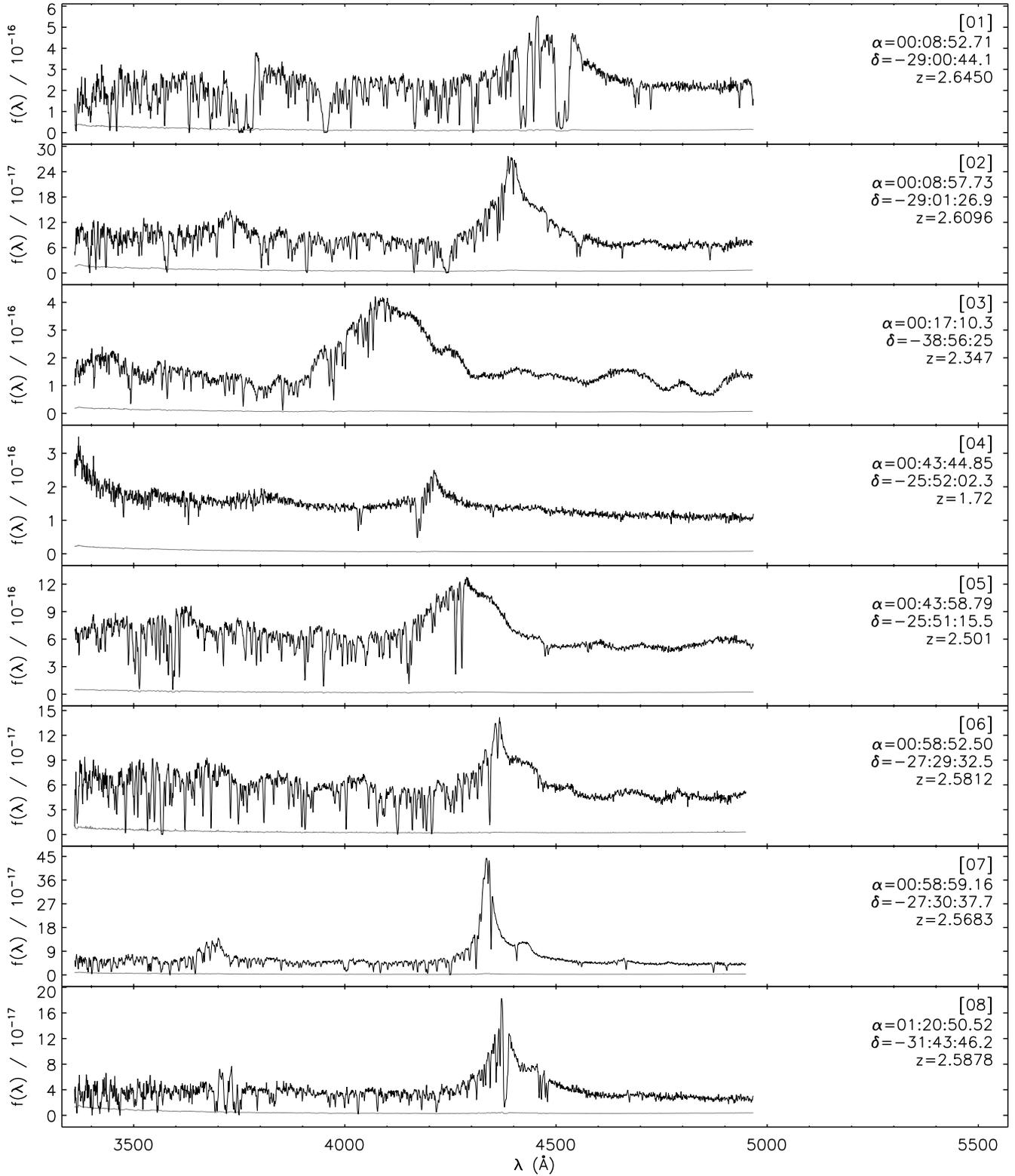}
\caption{Final combined, calibrated spectra for all observed
  objects are shown in black, plotted as flux (erg s$^{-1}$ cm$^{-2}$
  \AA$^{-1}$) vs. observed wavelength.  The corresponding 1 $\sigma$
  errors are shown in grey. Each spectrum is identified by an object
  number, the right
  ascension ($\alpha$) and declination ($\delta$) as J2000.0
  coordinates, and its redshift.}
\label{fig_spec}
\end{figure}

Figure~\ref{fig_spec} shows the resulting final, one-dimensional,
wavelength and flux calibrated spectra for \emph{all} objects observed
as part of this project.  These data, as well as the continuum fits
described in \S~\ref{sec_cont}, are also provided in
Table~\ref{tab_data} (the unabridged version is available in the electronic version of \apjs\ or upon request).
Incorrect published redshifts for one or both pair members 
resulted in the invalidation of four pairs.  In three other
cases, only one pair member was observed.  The observed QSOs with
$z>2.2$ are
detailed in Table~\ref{tbl_qsos} and comprise 35 pairs, one
triplet and six single lines of sight (an additional individual QSO
has $z=2.11$).  The latter are useful for
calculation of the autocorrelation function used in the AP test.
Although no formal maximum redshift was used and QSOs 
with redshifts as high as $z\gtrsim3$ are included, the decrease in
QSO number density beyond $z\sim2$ yields a mean redshift of
$\bar{z}=2.5$ and an effective upper limit of $z=2.8$.
As this paper was being
prepared for submission, we became aware of a complementary data set
\citep{2006cop370mnras1804} produced by a concurrent program using the
VLT telescopes.  Eleven pairs are shared in common with this sample.

\subsection{Resolution}

\begin{figure}[t] 
\begin{center}
\includegraphics[width=3.5in]{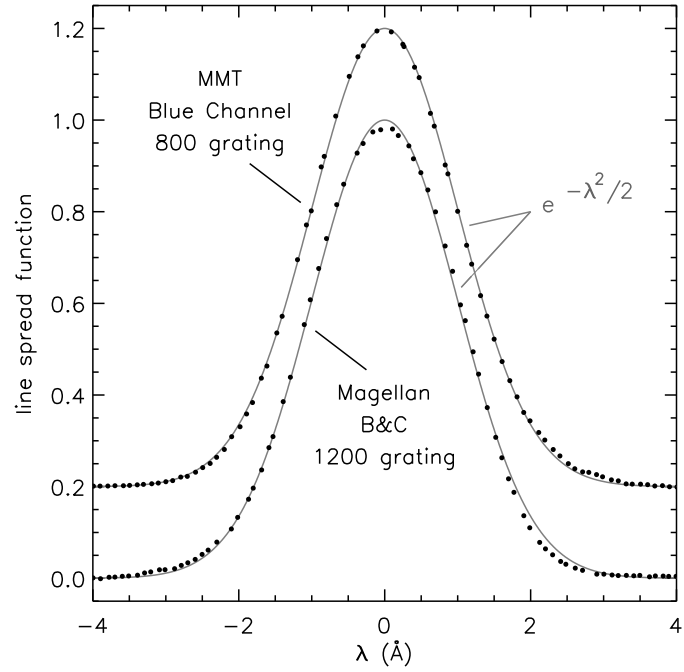}
\caption{Median LSFs for the B\&C and
  Blue Channel (shifted upward 0.2 for clarity) spectrographs are well
  parameterized as Gaussian.}
  \label{fig_lsf}
\end{center}
\end{figure}

The AP analysis presented in subsequent papers in this series
requires accurate knowledge of the spectral resolution of the data and assumes a Gaussian
line spread function (LSF).  Comparison lamp spectra taken immediately
before or after individual observations were used to
measure the former and verify the latter.  For each lamp spectrum, Gaussian
fits were made to sufficiently strong and unblended lines (see
Table~\ref{tbl_lines}), and the median width was taken to be the
spectral resolution of the corresponding QSO spectrum.  A composite
LSF was then created by aligning the lines from every
lamp after normalizing them by these fits (affecting the amplitude and
width, but not the shape).  Figure~\ref{fig_lsf} shows the composite median
for the B\&C and Blue Channel spectrographs.  In both cases, the LSF is indeed well parameterized as Gaussian.

The resulting spectral resolution for each QSO spectrum is included in
Table~\ref{tbl_qsos}.  In those cases in which multiple observations were
combined to form the final spectrum, the larger of the values is listed.
Generally, the variance in resolution is the result of normal
spectrograph focus degradation.  However, in rare cases, the measured
resolution were significantly broader than expected.  
Inspection of the observing logs indicates
that these instances occurred only at the MMT and always subsequent to
rotation of the grating carousel.  Where sufficient unaffected data
was available, these observations were not included in the final QSO
spectrum, resulting in slightly
lower S/N from what was originally anticipated.

\subsection{Data Quality}

The eighth column of Table~\ref{tbl_qsos}
lists the mean S/N per pixel in the unabsorbed portions of the
``pure'' \lya\ forest lying between the wings of the QSO's \lya\ and
Lyman$-\beta$/O~VI emission lines.  This is, of course, only a figure of
merit, as the S/N varies across each spectrum.  The target S/N in
this wavelength range was 10 per pixel, or greater.  The actual values
vary significantly due to factors such as deteriorated observing
conditions, fainter than expected QSOs, changed object priorities and
exclusion of exposures for various reasons.  In cases in which the target
S/N is met in at least a portion of the pure \lya\
forest (78 QSOs comprising 29 pairs and the one triplet), those
portions remain useful for the AP test.  For the transverse proximity
effect, only the S/N of the background QSO spectrum at the redshift of
the foreground QSO is important.  Of the relevant pairs, 17
have a mean S/N $>$ 5 in this region.

\subsection{Flat-Fielding Anomaly}

\begin{figure}[t] 
\begin{center}
\includegraphics[width=3.5in]{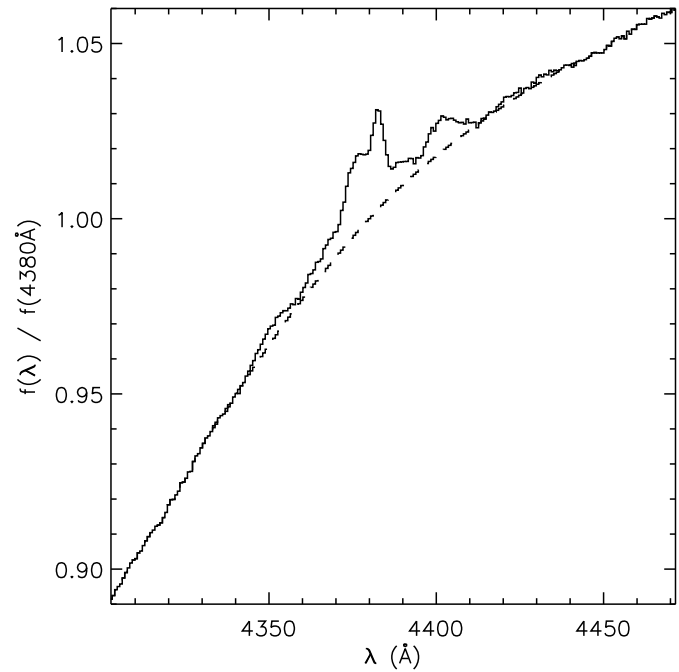}
\caption{Anomalous emission features present in spectra of the MMT Top
Box incandescent lamp, when observed with the 800 grating, prevent
reliable flat-fielding for the wavelength range 4330 \AA\ $< \lambda <$
4440 \AA.}
\label{fig_anomaly}
\end{center}
\end{figure}

Previously undocumented anomalous emission features were detected in
the dome flats taken at the MMT (see Fig.~\ref{fig_anomaly}).
These features persist throughout the 3 years of observations.
Although the origin is not known, we confirm that they occur at fixed
wavelengths and are unique to the 800 grating.  Such a spectral
response cannot be fit sufficiently without also partially removing
the pixel to pixel variations that the dome flats are designed to
characterize.  Our solution is to replace the affected region with a
smooth polynomial function.  As a result, the wavelength range falling
roughly between 4330 and 4440\AA\ in our MMT data has not been
flat-fielded (indicated by the last column in Table~\ref{tab_data}). 
This affects the \lya\ forest in only 
one QSO spectrum in this data set.

\section{Preamble To An AP Analysis}

\subsection{Continuum Fitting}\label{sec_cont}

The Alcock-Paczy\'{n}ski test relies on continuous flux statistics rather than
identification of discrete absorption lines.  However, all analysis of
spectral absorption requires knowledge of
the underlying, unabsorbed continuum.  This was estimated for the QSOs
in the sample using 
{\tt ANIMALS} \citep{2006pet132aj2046}, a continuum-fitting code based on the methodology employed
by the \emph{HST} QSO Absorption Line Key Project
\citep{1993bah87apjs1}.  Essentially, the spectrum is block-averaged and
fit with a cubic spline, which is then used to identify and mask
areas of absorption.  These steps are repeated until the fit
converges.  

Despite this relatively simple algorithm, obtaining a
credible continuum fit is generally a time-consuming and interactive
process. The smoothing scale is determined by the user and can be
adjusted across the spectrum as needed to compensate for varying
degrees of change in the slope of the continuum.  The threshold for
masking pixels that lie below the fit is a free parameter that
effectively raises or lowers the continuum.  If needed, the user can
manually mask or unmask pixels and lock some portions of the fit while
allowing others to be refined.  The continuum can even be set by hand in areas
where the fit is not satisfactory, such as boundaries between
different smoothing scales, heavily absorbed regions, or emission
features where the continuum changes rapidly.

\begin{figure}[t] 
\begin{center}
\includegraphics[width=7.25in]{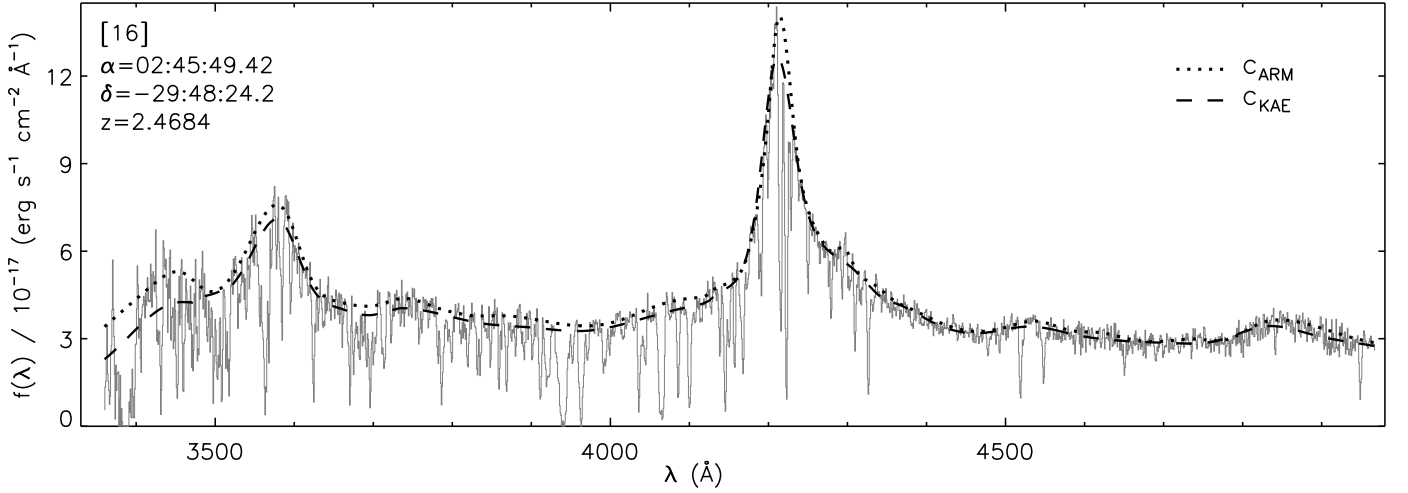}
\caption{Continua were fit independently by authors ARM ($c_{\rm ARM}$)
  and KAE ($c_{\,{\rm KAE}}$) in order to gauge the systematic error
  associated with the methodology for continuum estimation.  The
  example shown here is for a spectrum with mean ${\rm S/N} \approx 10$ per
  pixel in the
  \lya\ forest.}
\label{fig_contfit}
\end{center}
\end{figure}

In order to gauge the uncertainty in the rather subjective placement
of the continua, all of the QSOs were fit independently by authors ARM and
KAE.  Both fits are provided in Table~\ref{tab_data}, and are shown in
Figure~\ref{fig_contfit} for a typical spectrum with S/N~$\approx10$ per pixel in the
\lya\ forest.  One difference evident here (but also consistent
throughout the sample) is the lower continuum
placement of the latter ($c_{KAE}$) relative to the former ($c_{ARM}$).  Note also that
the continuum fits become increasingly unreliable with decreasing S/N
(\emph{e.g.}, S/N$<10$ at $\lambda < $3500\AA) and in close proximity to
 emission features (\emph{e.g.}, \lya\ at $\lambda\approx
4200$\AA\ and the \lyb/\osx\ blend at $\lambda\approx 3600$\AA).  In
some cases, the combination of \lya\ emission and strong absorption makes
estimation of the continuum impossible. For
this reason, and others, analysis of the \lya\ forest is generally
restricted to wavelengths less than $\sim3000$ km s$^{-1}$ blueward of
the \lya\ emission line.

\subsection{Auto and Cross-Correlation}

The autocorrelation ($\xi_{\parallel}$) is measured in the radial
direction and can be obtained independently from each QSO spectrum, although with
significant variance from one line of sight to another.  The
cross-correlation, on the otherhand, samples the transverse direction
and must be pieced together from pairs of QSOs at different separations:
 
\begin{eqnarray}
  \xi_{\parallel}(\Delta v) & = & 
  \left< 
  \left( \hat{f}(v)\,/\,\langle \hat{f} \rangle - 1 \right)
  \left( \hat{f}(v+\Delta v)\,/\,\langle \hat{f} \rangle - 1\right) 
  \right>
\label{eq_auto} 
\\
  \xi_{\perp}\left(\Delta v\right) & = & 
  \left<
  \left( \hat{f}(v)\,/\,\langle \hat{f} \rangle - 1 \right)
  \left( \hat{f}_{\Delta v}(v)\,/\,\langle \hat{f} \rangle - 1 \right)
  \right>
  \label{eq_cross}
\end{eqnarray}

\begin{figure}[t] 
\begin{center}
\includegraphics[width=3.5in]{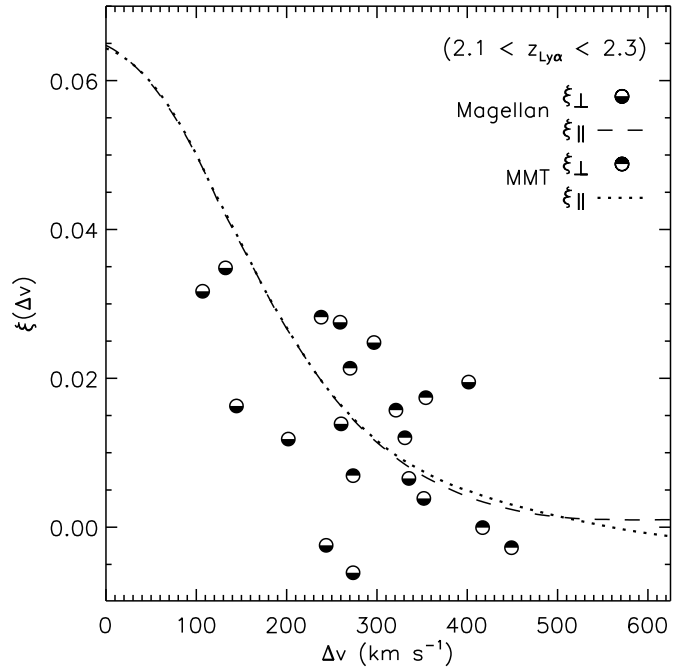}
\caption{Autocorrelation ($\xi_\parallel$) and cross-correlation
  ($\xi_\perp$) measurements for the redshift range
  $2.1 < z_{Ly\alpha} < 2.3$, assuming $\Omega_m = 0.268$ and 
  $\Omega_{\Lambda} = 0.732$.  Anisotropies due to redshift-space
  distortions and 
  spectral resolution must be accounted for prior to an AP analysis.
}\label{fig_correlation}
\end{center}
\end{figure}

\noindent In equations~(\ref{eq_auto}) and (\ref{eq_cross}), $\hat{f}$ and
$\hat{f}_{\Delta v}$ correspond to 
lines of sight separated on the sky by $\Delta v$, where $\hat{f}$ is the flux
divided by the continuum. Figure~\ref{fig_correlation} shows
correlation measurements for one redshift bin ($2.1 < z_{Ly\alpha} <
2.3$), assuming the $c_{KAE}$ continua, the observationally determined
value of $\langle \hat{f} \rangle$ from \citet{1993pre414apj64},
$\Omega_m = 0.268$, and $\Omega_\Lambda = 0.732$.  The agreement
between the autocorrelation and cross-correlation for the currently
preferred cosmological model \citep{2007spe}
is, in part, a coincidence.  Two sources
of anisotropy in the \lya\ flux correlation function must be accounted
for before a reliable AP analysis can be carried out.  First, the line
spread function of the spectrograph smooths the spectra along the line
of sight and therefore affects autocorrelation and
cross-correlation measurements differently.  Second, peculiar
velocities caused by the expansion of the universe, gravitational
collapse, and thermal broadening make the correlation function
anisotropic in redshift (observed) space \citep{1987kai227mnras1}.  Modelling these effects
with cosmological hydrodynamic simulations is the subject of Paper II
in this series.

\section{QSO Catalog Corrections}\label{errata}

In the course of our observations, five QSOs were found to have
incorrect published redshifts and two additional objects turned 
out not to be QSOs at all.  These errors were the result of 
misidentification based on inferior data available 
at the time, with one exception.  \citet{2006cop370mnras1804}
and this 
paper present spectra for the same QSO, 2QZ~J102827.1$-$013641, which
are clearly different and yield disparate redshifts ($z=2.393$ and 
1.609, respectively).  The original redshift ($z=2.31$) obtained by 
\citet{2004cro349mnras1397}
disagrees with both follow-up studies,
but appears to result from the \cfr\ emission line being
mistakenly identified as \ly.  The fact that their discovery spectrum 
is consistent with being a much noisier version of the spectrum 
presented in this paper, leads us to conclude that our redshift is 
correct and that the \citet{2006cop370mnras1804} spectrum is
for a QSO at different coordinates.

Table~\ref{tbl_newz} details these errors
and provides corrections based on our observations.
The fourth column lists the mean and standard deviation of redshifts
determined from the emission lines recorded in the fifth column.
Gaussian curves were fit to each line, and the corresponding central
wavelengths were compared to the observationally determined rest
wavelengths reported by \citet{2001van122aj549}.

\section{Summary}\label{summary}

We have carried out a 3 year observational campaign to obtain
optical spectroscopy for pairs of previously known
QSOs with $z>2.2$, $\Delta z < 0.25$ and separations less than 5
arcminutes$^2$:

1) We present 86 spectra comprising 35 QSO pairs, one triplet, 
11 individual QSOs (four of which have $z\lesssim1.7$), and two objects
previously misidentified as QSOs. 

2) Seven previously catalogued QSOs were found to have incorrect
published identifications and/or redshifts, for which we provide
revised redshift values. 

3) We note one aspect of observing with the MMT Blue Channel
spectrograph that may be relevant for future users.  The otherwise
smooth spectral response of the incandescent lamp in the Top Box
exhibits anomalous emission features at $\lambda\approx4380$\AA\
when the 800 grating is used.

4) We have created a new spectroscopic dataset of 78 QSOs with
sufficient data quality for an Alcock-Paczy\'{n}ski
measurement of $\Omega_{\Lambda}$; 29 pairs and one triplet for measuring the
cross-correlation of transmitted flux in the \lya\ forest,
and 17 additional individual QSOs for measuring the
autocorrelation. 

5) In addition, 17 of the QSO pairs are both unassociated and
have sufficient data quality for an investigation of the transverse
proximity effect.

\acknowledgements

We gratefully acknowledge the operators and staff at the MMT and
Magellan telescopes for their assistance and expertise that made this
observational program possible, Daniel Christlein for helping observe
during the winter holiday in 2002, and John Moustakas for countless
conversations regarding the {\tt iSPEC2d} reduction package.
This research has made use of the NASA/IPAC Extragalactic Database (NED) 
which is operated by the Jet Propulsion Laboratory, California Institute 
of Technology, under contract with the National Aeronautics and Space Administration.

%%!%% Facilities added
{\it Facilities:} \facility{MMT (Blue Channel spectrograph)}, 
\facility{Magellan:Baade (Boller and Chivens spectrograph)}, 
\facility{Magellan:Clay (Boller and Chivens spectrograph)}

\clearpage
\clearpage
\begin{center}
\includegraphics[scale=0.96]{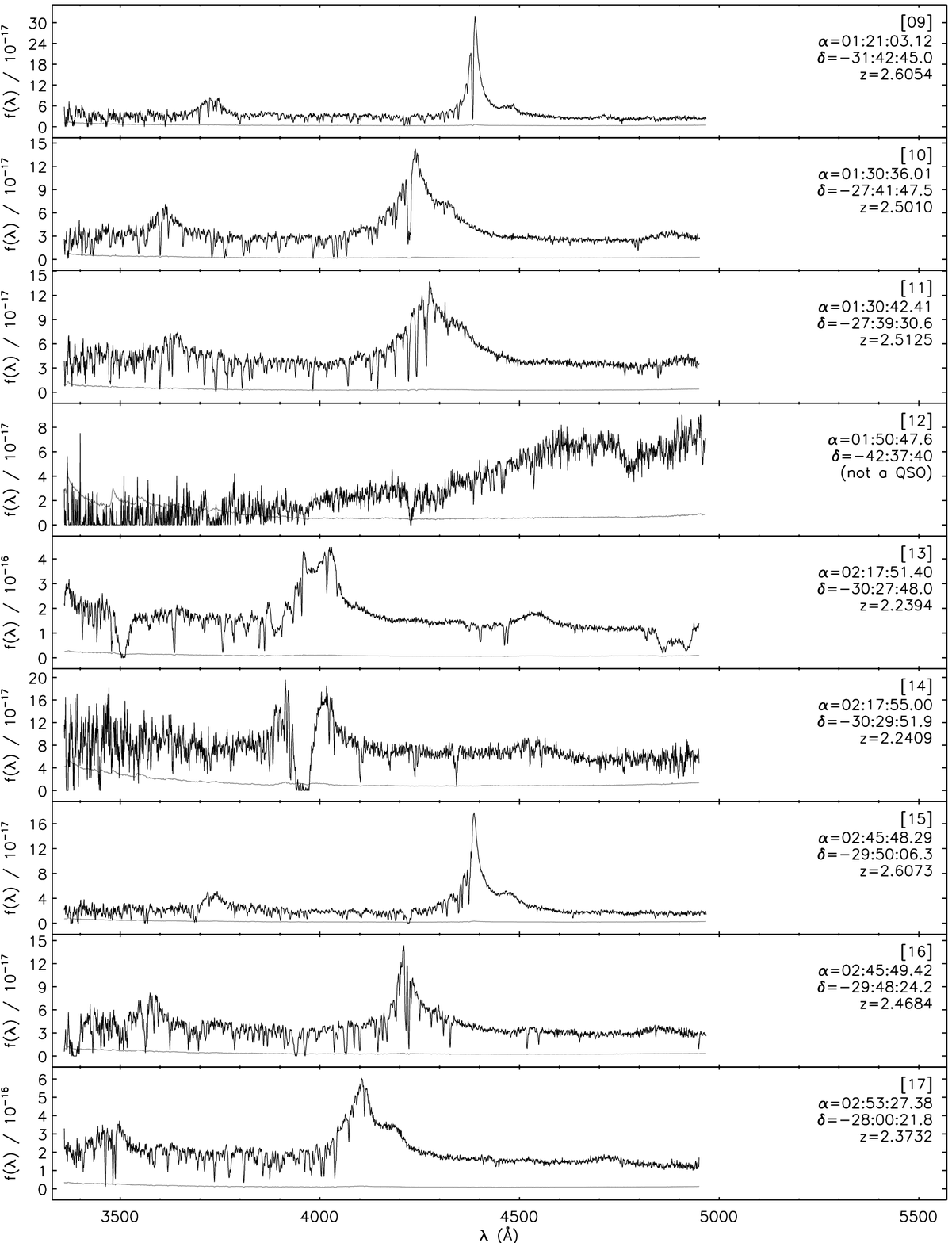}\\[5mm]
\centerline{Fig. 1. --- continued...}
\clearpage
\includegraphics[scale=0.96]{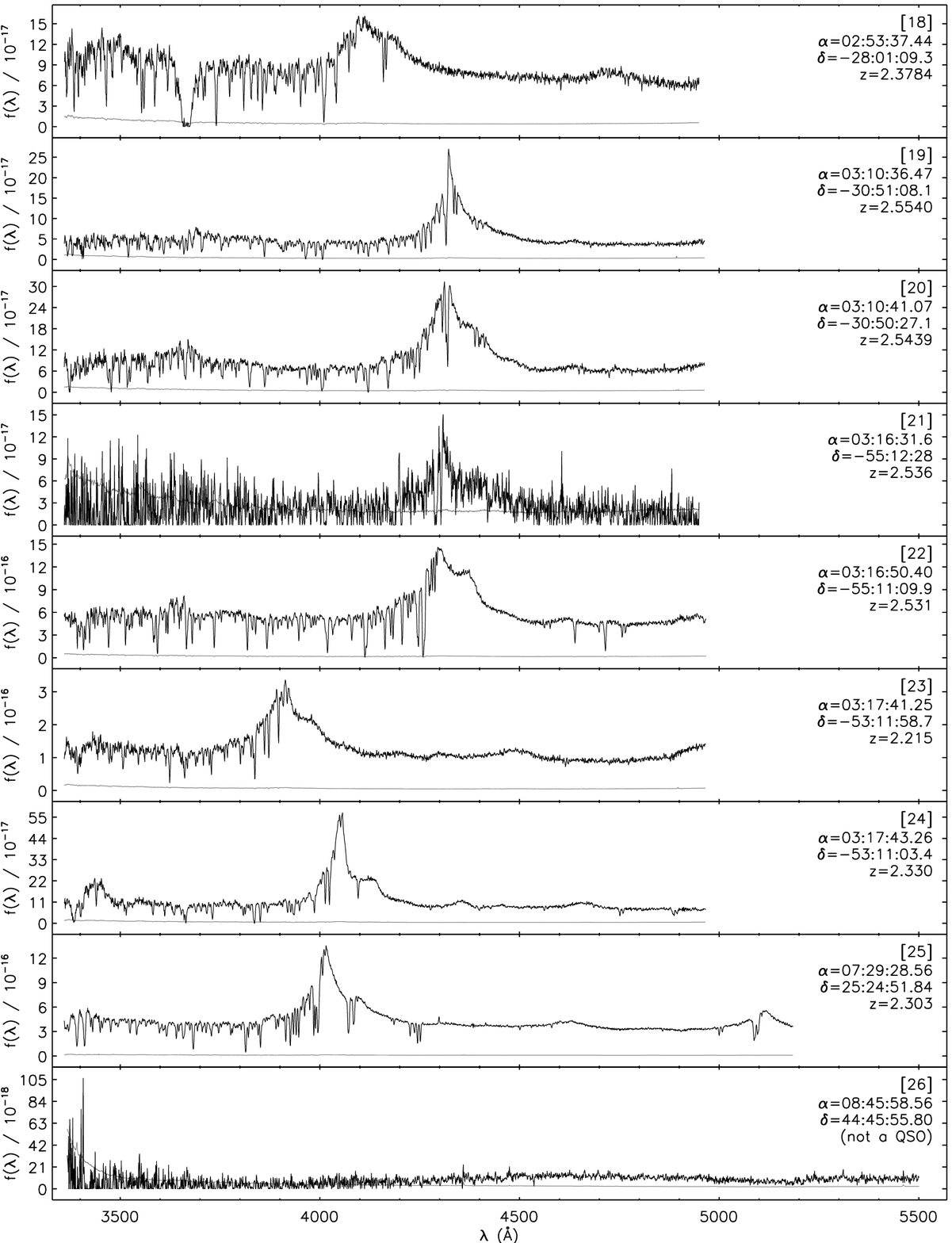}\\[5mm]
\centerline{Fig. 1. --- continued...}
\clearpage
\includegraphics[scale=0.96]{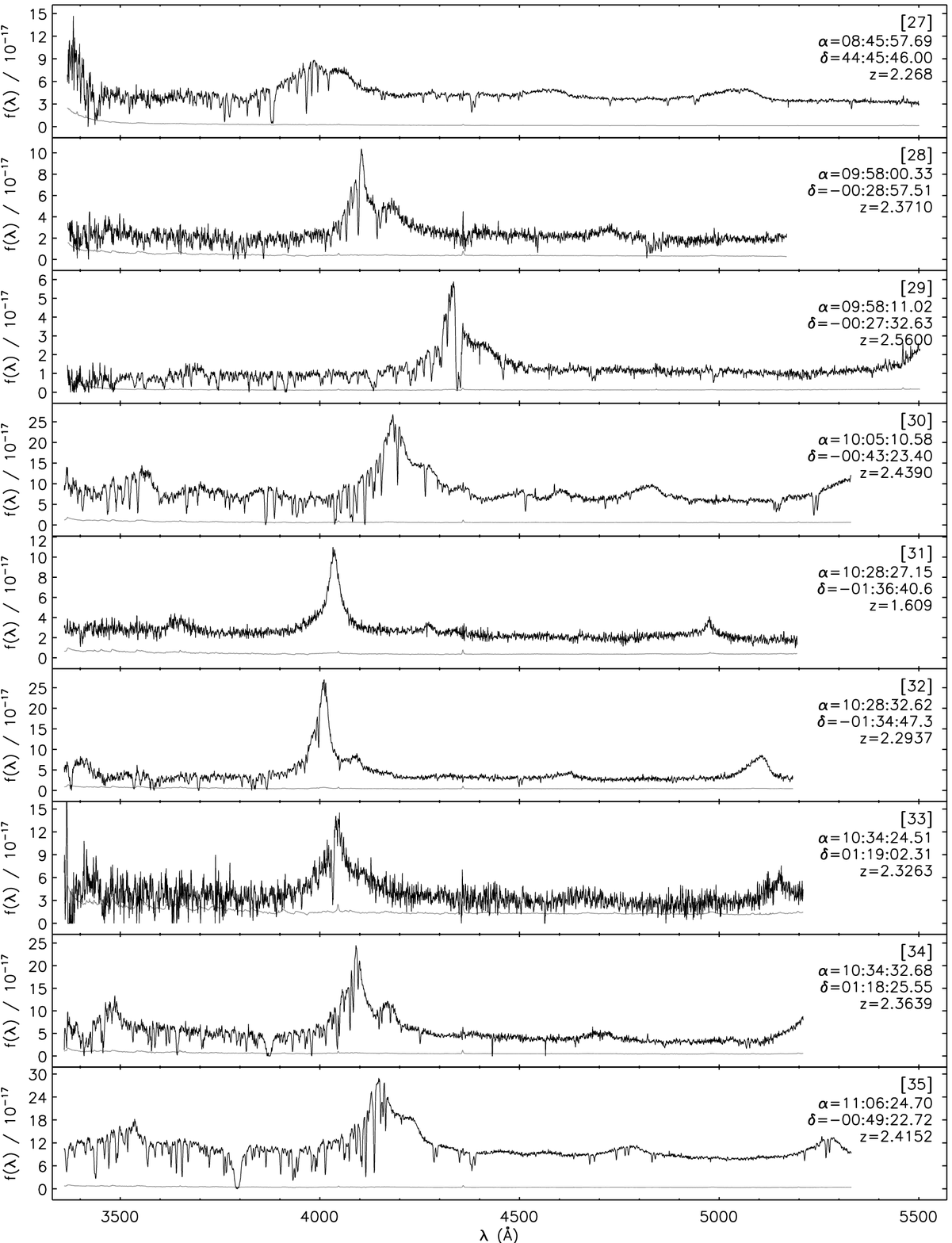}\\[5mm]
\centerline{Fig. 1. --- continued...}
\clearpage
\includegraphics[scale=0.96]{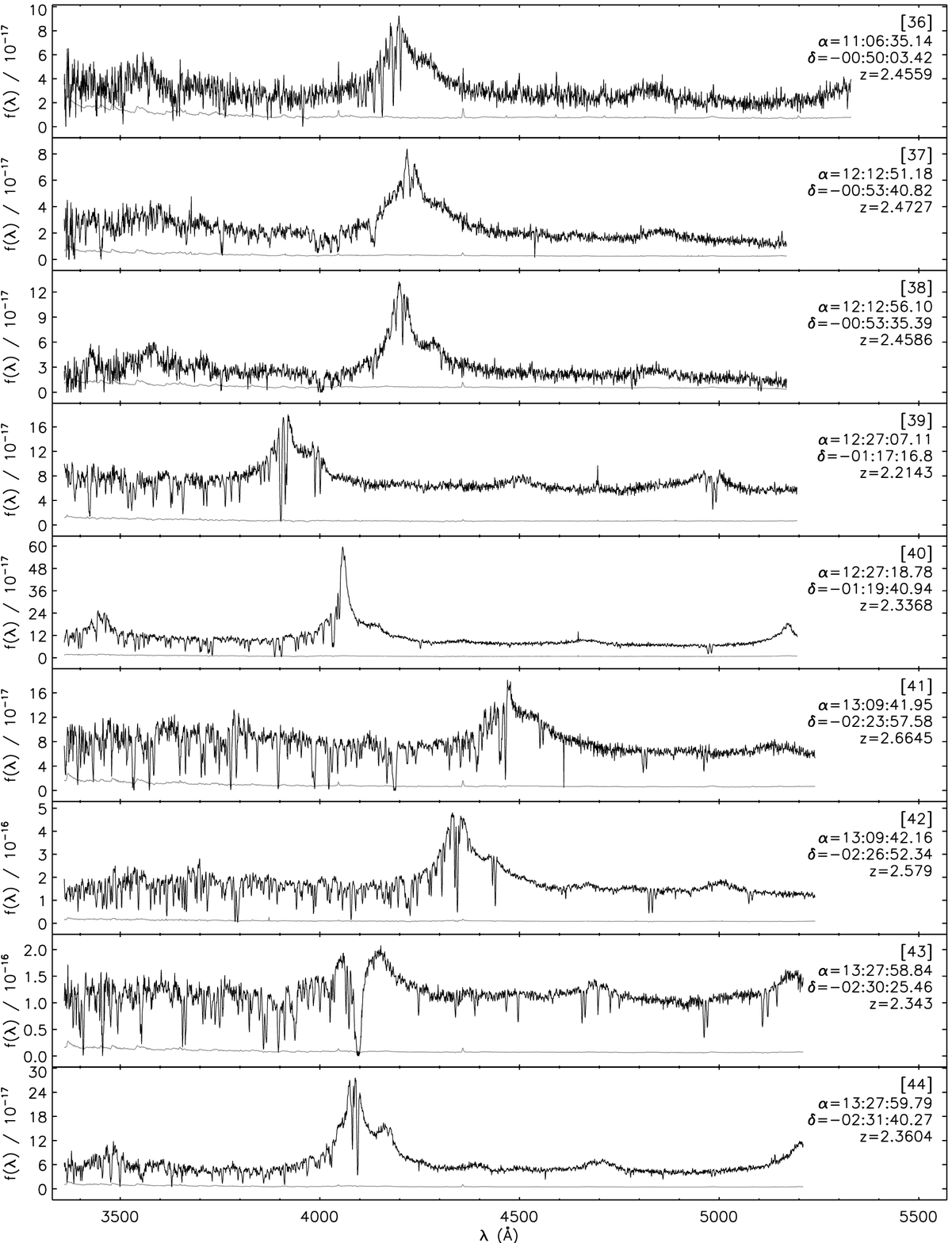}\\[5mm]
\centerline{Fig. 1. --- continued...}
\clearpage
\includegraphics[scale=0.96]{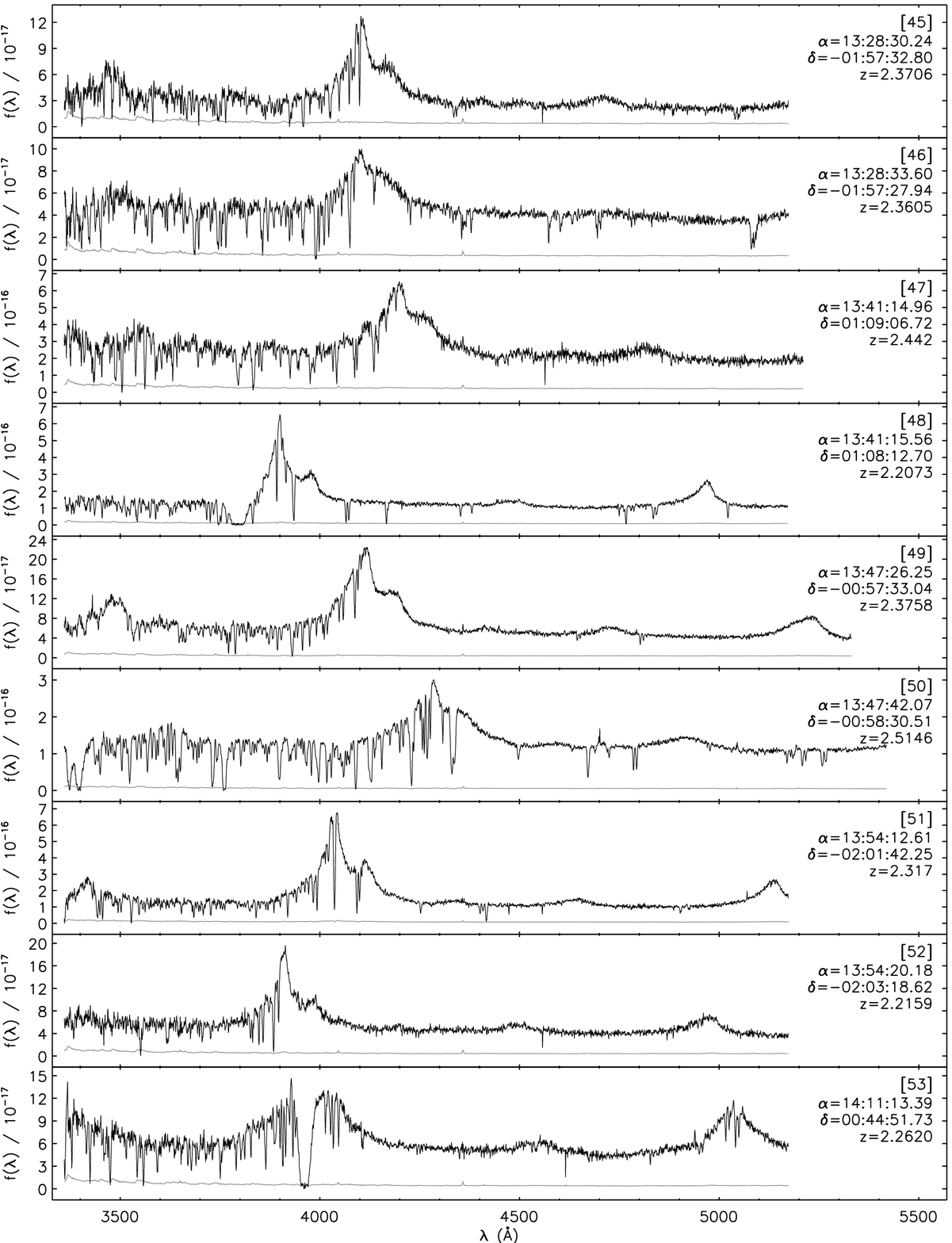}\\[5mm]
\centerline{Fig. 1. --- continued...}
\clearpage
\includegraphics[scale=0.96]{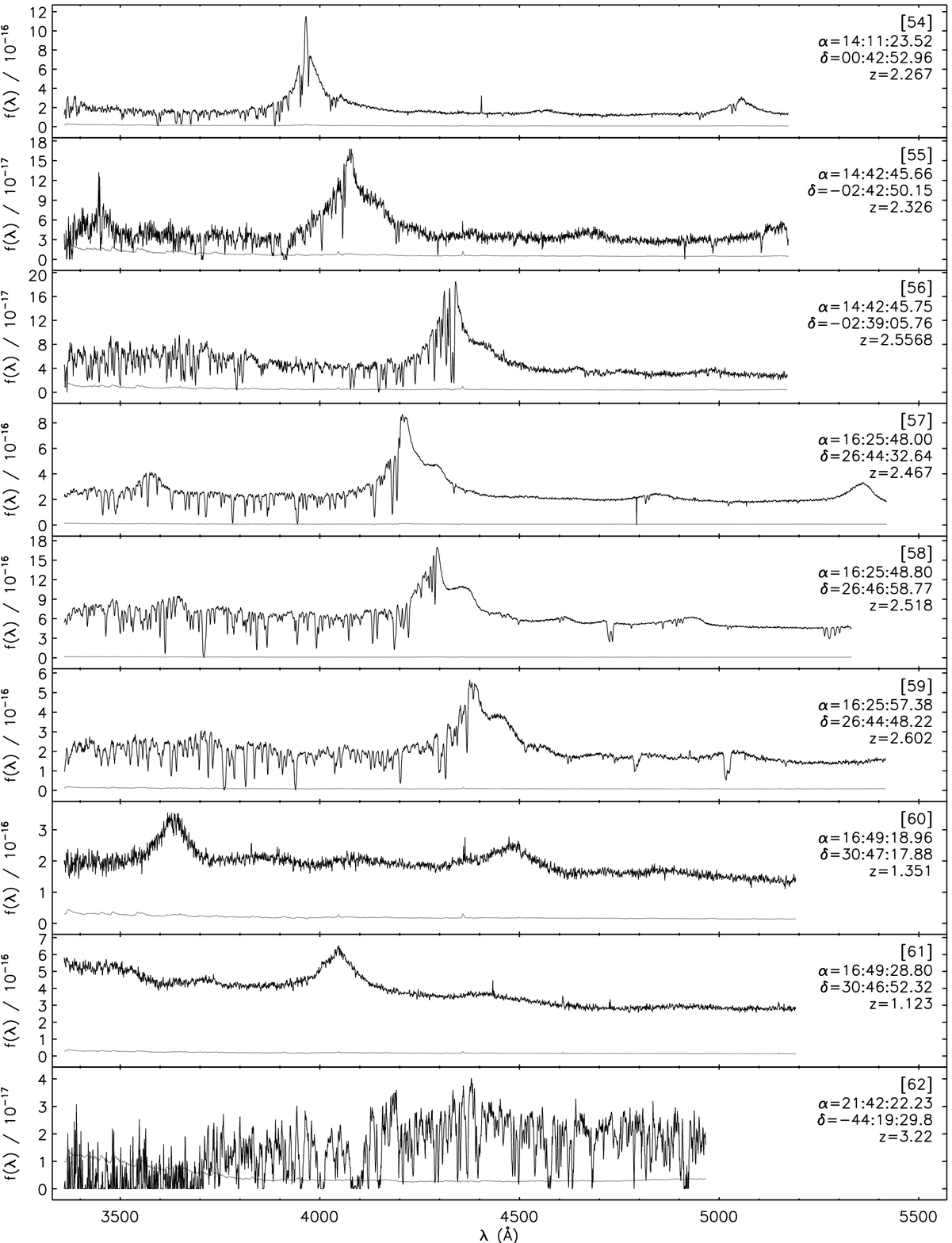}\\[5mm]
\centerline{Fig. 1. --- continued...}
\clearpage
\includegraphics[scale=0.96]{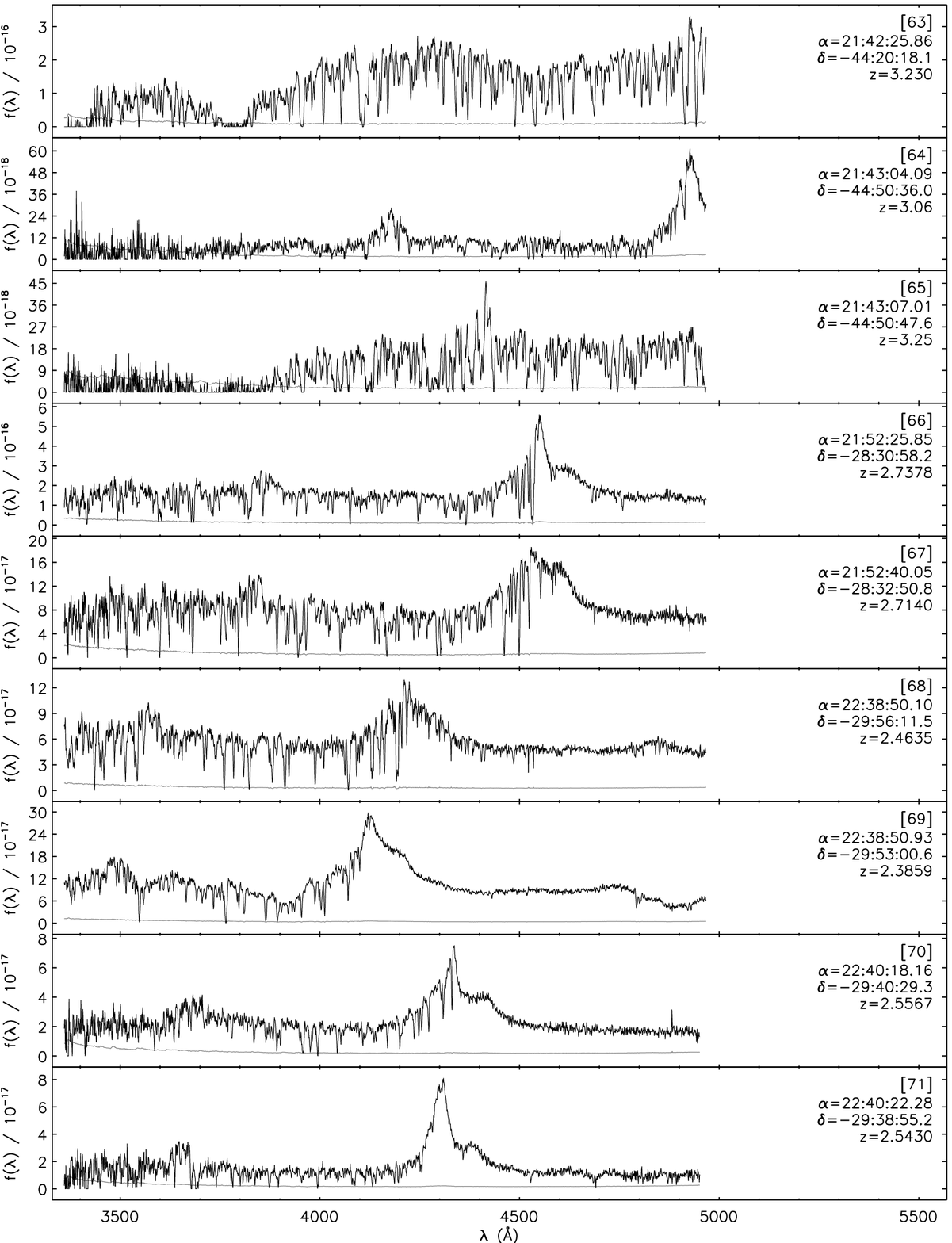}\\[5mm]
\centerline{Fig. 1. --- continued...}
\clearpage
\includegraphics[scale=0.96]{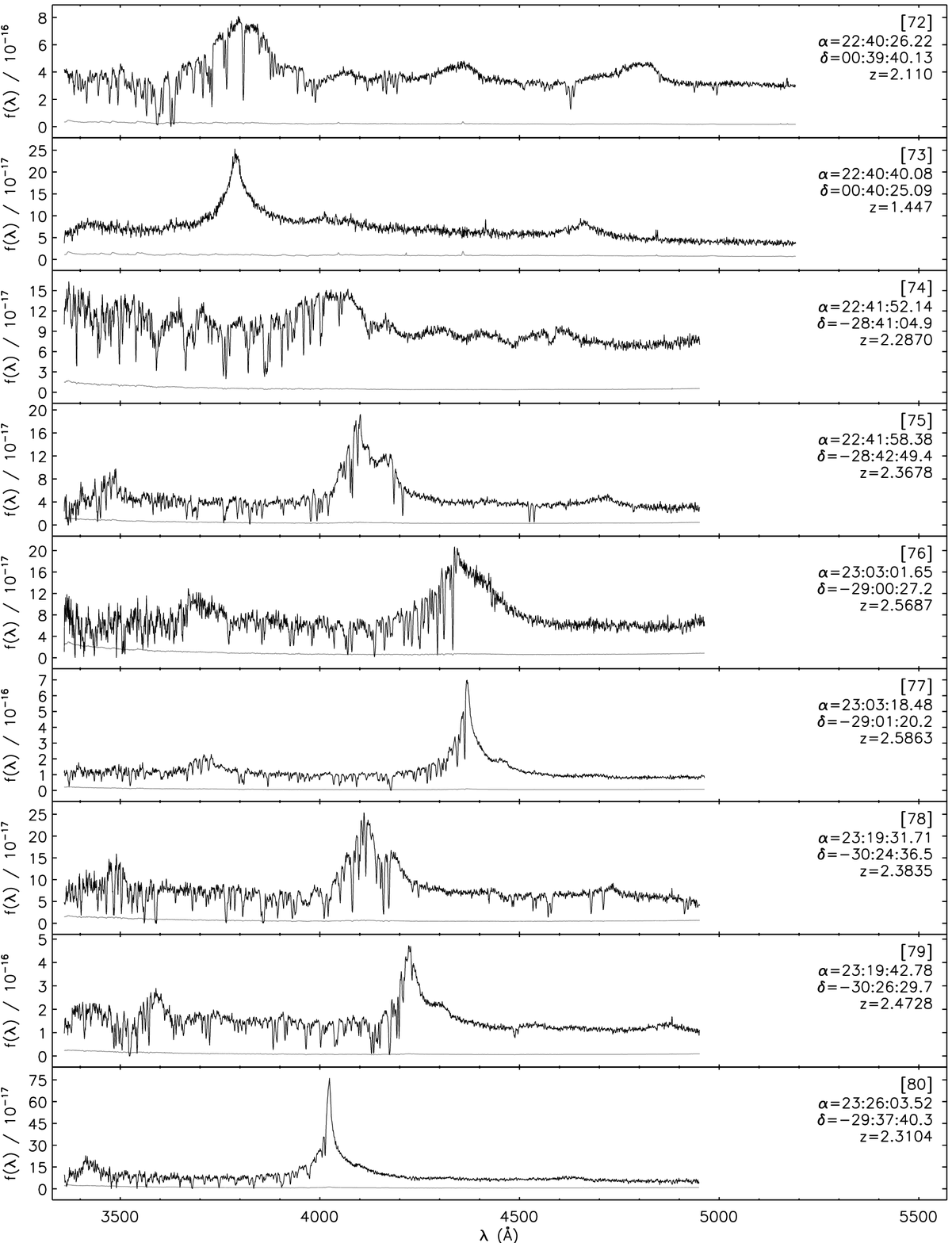}\\[5mm]
\centerline{Fig. 1. --- continued...}
\clearpage
\includegraphics[scale=0.96]{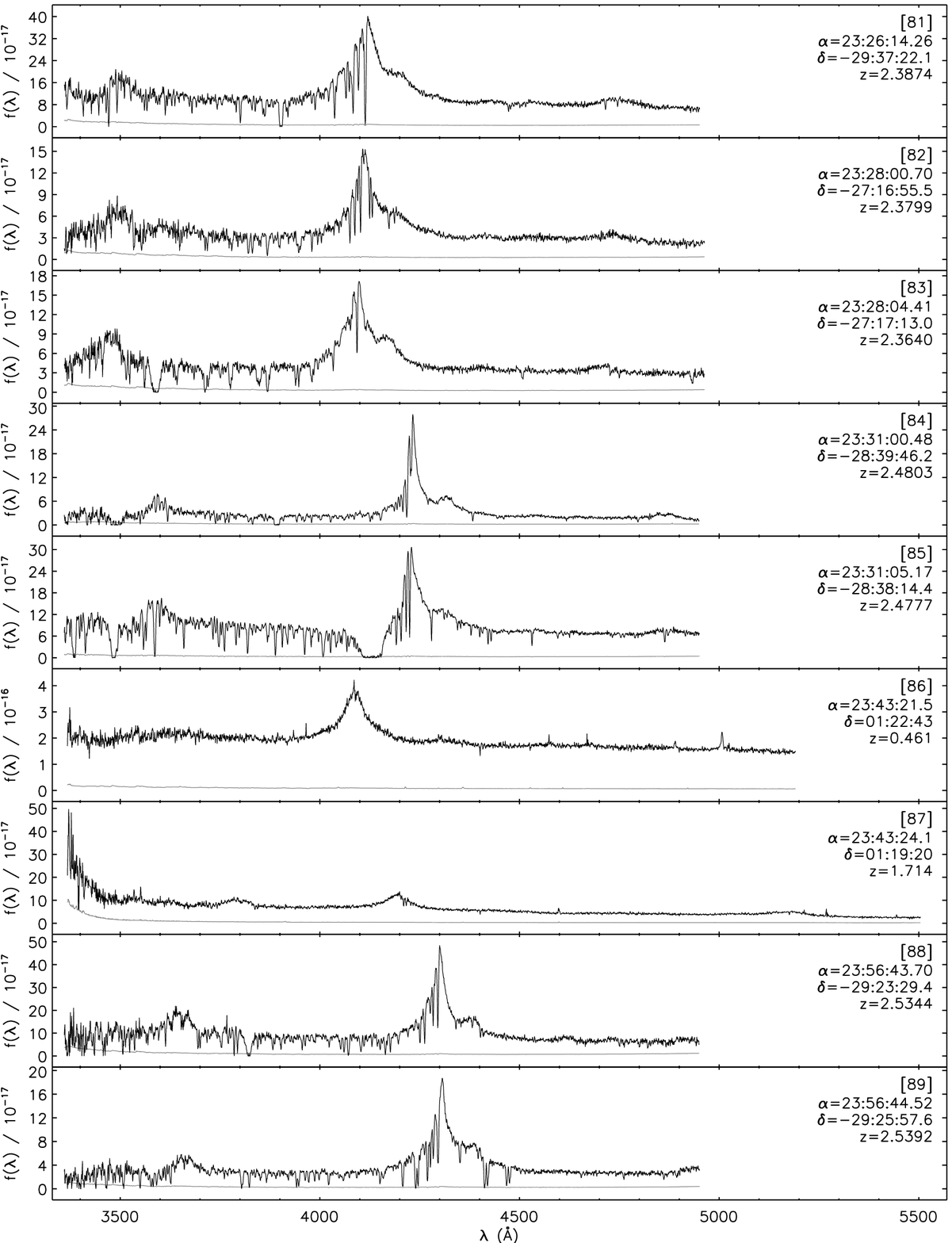}\\[5mm]
\centerline{Fig. 1. --- continued...}
\end{center}
\clearpage

\clearpage
\tabletypesize{\footnotesize}
%%%%%%%%%%%%%%%%%%%%%%%%%%%%%%%%%%%%%%%%%%%%%%%%%%%%%%%%%%%
%%                                                       %%
%% THIS TEX FILE WAS CREATED BY AN IDL SCRIPT AND SHOULD %%
%%  NOT BE MODIFIED - MAKE CHANGES TO SCRIPT INSTEAD...  %%
%%                                                       %%
%%%%%%%%%%%%%%%%%%%%%%%%%%%%%%%%%%%%%%%%%%%%%%%%%%%%%%%%%%%

\begin{deluxetable}{c c c c c c c}
\tablecolumns{7}
\tabletypesize{\scriptsize}
\tablewidth{6.5in}
\tablecaption{Spectroscopic Data. \emph{(abridged)}\label{tab_data}}
\tablehead{
\colhead{} & \colhead{$\lambda$} & \colhead{$f$} &
  \colhead{$\sigma_f$} &
  \colhead{$c_{\rm ARM}$} &
  \colhead{$c_{\,{\rm KAE}}$} &
  \colhead{} \\
%\cline{3-6}\\[-0.25cm]
\colhead{Number} & \colhead{(\AA)} &
\colhead{(erg s$^{-1}$ cm$^{-2}$ \AA$^{-1}$)} & 
\colhead{(erg s$^{-1}$ cm$^{-2}$ \AA$^{-1}$)} & 
\colhead{(erg s$^{-1}$ cm$^{-2}$ \AA$^{-1}$)} & 
\colhead{(erg s$^{-1}$ cm$^{-2}$ \AA$^{-1}$)} & 
\colhead{Flag\tablenotemark{a}}\\[-0.25cm]}
\startdata
01 & 3360.00 & 1.370E-16 & 3.753E-17 & 1.864E-16 & 1.567E-16 & 0\\
01 & 3360.80 & 1.242E-16 & 3.540E-17 & 1.872E-16 & 1.573E-16 & 0\\
01 & 3361.60 & 6.800E-17 & 3.374E-17 & 1.880E-16 & 1.579E-16 & 0\\
$\downarrow$ & $\downarrow$ & $\downarrow$ & $\downarrow$ & $\downarrow$ & $\downarrow$ & $\downarrow$\\
01 & 4967.20 & 1.599E-16 & 1.492E-17 & 2.470E-16 & 2.490E-16 & 0\\
02 & 3360.00 & 5.036E-17 & 1.562E-17 & 9.525E-17 & 9.457E-17 & 0\\
02 & 3360.80 & 4.205E-17 & 1.557E-17 & 9.544E-17 & 9.455E-17 & 0\\
02 & 3361.60 & 4.930E-17 & 1.594E-17 & 9.564E-17 & 9.449E-17 & 0\\
$\downarrow$ & $\downarrow$ & $\downarrow$ & $\downarrow$ & $\downarrow$ & $\downarrow$ & $\downarrow$\\
\enddata
\tablecomments{The columns (left to right) are: object number, wavelength,
  flux, 1-$\sigma$ flux uncertainty, 
  continuum fit by ARM, continuum fit by KAE,
  and a descriptive flag.  [\emph{The unabridged version of this table 
      can be found in the electronic edition of \apjs\ or upon request.}]}
\tablenotetext{a}{Non-zero flags denote irregularities in the data:
  1 -- not available, 2 -- not flat-fielded}
\end{deluxetable}

\tabletypesize{\scriptsize}
%\clearpage
\LongTables
%\begin{landscape}
\begin{deluxetable}{c c l l l l c@{ }c@{ }c@{ }l c c l}
\tablecolumns{13}
\tablewidth{0pc}
\tablecaption{Sample QSOs.\label{tbl_qsos}}
\tablehead{
\colhead{} &
\colhead{$\theta$} & \colhead{$\alpha$} & \colhead{$\delta$} & \colhead{} & \colhead{} & 
\multicolumn{4}{c}{Observations} & \colhead{FWHM} & 
\colhead{} & \colhead{}\\
\colhead{Number} & 
\colhead{(arcmin)} & \colhead{(J2000.0)} & \colhead{(J2000.0)} & \colhead{$z$} & \colhead{$M$ (various\tablenotemark{a})} & 
\multicolumn{4}{c}{(UT)} & \colhead{(\AA)} &
\colhead{$\langle$S/N$\rangle$} & \colhead{Alias(es)}}

\startdata

\multicolumn{13}{c}{QSO Pairs}\\[.1cm]
\hline\\[-0.25cm]
01 & 1.31 & 00:08:52.71\tablenotemark{1}  & $-$29:00:44.1  & 2.6450\tablenotemark{1}   & 19.115\tablenotemark{1} & 2002&OCT&30&CLAY     & 2.56 &    21 &     \object{2QZ J000852.7$-$290044}\tablenotemark{1}\\
02 &      & 00:08:57.73\tablenotemark{1}  & $-$29:01:26.9  & 2.6096\tablenotemark{1}   & 19.846\tablenotemark{1} & 2002&OCT&30&CLAY     & 2.55 &    16 &     \object{2QZ J000857.7$-$290126}\tablenotemark{1}\\
\hline\\[-0.25cm]		     						  					                   
06 & 1.83 & 00:58:52.50\tablenotemark{1}  & $-$27:29:32.5  & 2.5812\tablenotemark{1}   & 20.086\tablenotemark{1} & 2002&NOV&02&CLAY     & 2.57 &    23 &                             SGP4:27\tablenotemark{2}\\
   &      &                               &                &                           &                         & 2003&SEP&30&BAADE    &      &       &     \object{2QZ J005852.4$-$272933}\tablenotemark{1}\\
07 &      & 00:58:59.16\tablenotemark{1}  & $-$27:30:37.7  & 2.5683\tablenotemark{1}   & 19.904\tablenotemark{1} & 2002&NOV&02&CLAY     & 2.53 &    19 &     \object{2QZ J005859.1$-$273038}\tablenotemark{1}\\
   &      &                               &                &                           &                         & 2003&SEP&30&BAADE    &      &       &                                                     \\
\hline\\[-0.25cm]		     						  					                            
08 & 2.87 & 01:20:50.52\tablenotemark{1}  & $-$31:43:46.2  & 2.5878\tablenotemark{1}   & 20.244\tablenotemark{1} & 2002&NOV&02&CLAY     & 2.50 &    11 &     \object{2QZ J012050.5$-$314346}\tablenotemark{1}\\
09 &      & 01:21:03.12\tablenotemark{1}  & $-$31:42:45.0  & 2.6054\tablenotemark{1}   & 20.635\tablenotemark{1} & 2002&NOV&02&CLAY     & 2.56 &    10 &     \object{2QZ J012103.1$-$314245}\tablenotemark{1}\\
\hline\\[-0.25cm]		     						  					       	                    
10 & 2.69 & 01:30:36.01\tablenotemark{1}  & $-$27:41:47.5  & 2.5010\tablenotemark{1}   & 20.731\tablenotemark{1} & 2003&SEP&29&BAADE    & 2.51 &    14 &     \object{2QZ J013035.9$-$274148}\tablenotemark{1}\\
11 &      & 01:30:42.41\tablenotemark{1}  & $-$27:39:30.6  & 2.5125\tablenotemark{1}   & 20.831\tablenotemark{1} & 2003&SEP&29&BAADE    & 2.51 &    14 &     \object{2QZ J013042.3$-$273931}\tablenotemark{1}\\
\hline\\[-0.25cm]		     						  					       	      	         
13 & 2.21 & 02:17:51.40\tablenotemark{1}  & $-$30:27:48.0  & 2.2394\tablenotemark{1}   & 19.340\tablenotemark{1} & 2003&SEP&28&BAADE    & 2.50 &    13 &     \object{2QZ J021751.3$-$302748}\tablenotemark{1}\\
14 &      & 02:17:55.00\tablenotemark{1}  & $-$30:29:51.9  & 2.2409\tablenotemark{1}   & 20.682\tablenotemark{1} & 2003&SEP&28,29&BAADE & 2.51 & \phn4 &     \object{2QZ J021754.9$-$302952}\tablenotemark{1}\\
   &      &                               &                &                           &                         & 2003&OCT&01&BAADE    &      &       &                                                     \\
\hline\\[-0.25cm]		     						  					       	      
15 & 1.72 & 02:45:48.29\tablenotemark{1}  & $-$29:50:06.3  & 2.6073\tablenotemark{1}   & 20.716\tablenotemark{1} & 2002&NOV&01&CLAY     & 2.51 &    10 &     \object{2QZ J024548.2$-$295006}\tablenotemark{1}\\
16 &      & 02:45:49.42\tablenotemark{1}  & $-$29:48:24.2  & 2.4684\tablenotemark{1}   & 20.626\tablenotemark{1} & 2002&NOV&01&CLAY     & 2.55 &    11 &     \object{2QZ J024549.4$-$294824}\tablenotemark{1}\\
\hline\\[-0.25cm]		     						  					       	           
17 & 2.36 & 02:53:27.38\tablenotemark{1}  & $-$28:00:21.8  & 2.3732\tablenotemark{1}   & 18.751\tablenotemark{1} & 2003&SEP&28&BAADE    & 2.50 &    14 &     \object{2QZ J025327.3$-$280022}\tablenotemark{1}\\
18 &      & 02:53:37.44\tablenotemark{1}  & $-$28:01:09.3  & 2.3784\tablenotemark{1}   & 20.309\tablenotemark{1} & 2003&SEP&28&BAADE    & 2.50 &    15 &     \object{2QZ J025337.4$-$280109}\tablenotemark{1}\\
\hline\\[-0.25cm]		     		      				  				       		           
19 & 1.20 & 03:10:36.47\tablenotemark{1}  & $-$30:51:08.1  & 2.5540\tablenotemark{1}   & 20.350\tablenotemark{1} & 2002&OCT&31&CLAY     & 2.63 &    14 &     \object{2QZ J031036.4$-$305108}\tablenotemark{1}\\
   &      &                               &                &                           &                         & 2002&NOV&02&CLAY     &      &       &                                                     \\
20 &      & 03:10:41.07\tablenotemark{1}  & $-$30:50:27.1  & 2.5439\tablenotemark{1}   & 19.549\tablenotemark{1} & 2002&OCT&31&CLAY     & 2.63 &    14 &     \object{2QZ J031041.0$-$305027}\tablenotemark{1}\\
   &      &                               &                &                           &                         & 2002&NOV&01,02&CLAY  &      &       &                                                     \\
\hline\\[-0.25cm]		     						  						      
21 & 2.98 & 03:16:31.6\tablenotemark{3}   & $-$55:12:28    & 2.536\tablenotemark{3}    & 21.40\tablenotemark{4}  & 2003&OCT&01&BAADE    & 2.50 & \phn1 &                    \object{MZZ4959}\tablenotemark{3}\\
22 &      & 03:16:50.40\tablenotemark{5}  & $-$55:11:09.9  & 2.531\tablenotemark{3}    & 18.04\tablenotemark{4}  & 2002&NOV&02&CLAY     & 2.51 &    27 &                    \object{MZZ4875}\tablenotemark{3}\\
   &      &                               &                &                           &                         &     &   &  &         &      &       &                               CT425\tablenotemark{5}\\
\hline\\[-0.25cm]		     						  					       	           
23 & 0.97 & 03:17:41.25\tablenotemark{5}  & $-$53:11:58.7  & 2.215\tablenotemark{6}    & 19.1\tablenotemark{5}   & 2002&OCT&30&CLAY     & 2.56 &    12 &                      \object{CT426}\tablenotemark{5}\\
   &      &                               &                &                           &                         & 2003&NOV&02&CLAY     &      &       &                                                     \\ 
24 &      & 03:17:43.26\tablenotemark{5}  & $-$53:11:03.4  & 2.330\tablenotemark{5}    & 19.1\tablenotemark{5}   & 2002&OCT&30&CLAY     & 2.56 &    13 &                      \object{CT427}\tablenotemark{5}\\
\hline\\[-0.25cm]		     						  					       	      
28 & 3.05 & 09:58:00.33\tablenotemark{7}  & $-$00:28:57.51 & 2.3710\tablenotemark{1}   & 20.534\tablenotemark{7} & 2003&JAN&06&MMT      & 3.57 & \phn4 &     \object{2QZ J095800.2$-$002858}\tablenotemark{1}\\
   &      &                               &                &                           &                         & 2003&MAR&08&MMT      &      &       &                                                     \\
29 &      & 09:58:11.02\tablenotemark{7}  & $-$00:27:32.63 & 2.5600\tablenotemark{1}   & 20.012\tablenotemark{7} & 2003&JAN&06&MMT      & 3.42 & \phn5 &     \object{2QZ J095810.9$-$002733}\tablenotemark{1}\\
\hline\\[-0.25cm]		     						  					       	      	         
33 & 2.13 & 10:34:24.51\tablenotemark{7}  & $+$01:19:02.31 & 2.3263\tablenotemark{1}   & 20.340\tablenotemark{7} & 2004&MAR&28&MMT      & 2.91 & \phn2 &       \object{2QZ J103424.4+011901}\tablenotemark{1}\\
34 &      & 10:34:32.68\tablenotemark{7}  & $+$01:18:25.55 & 2.3639\tablenotemark{1}   & 19.851\tablenotemark{7} & 2003&MAR&27&MMT      & 2.60 & \phn8 &       \object{2QZ J103432.6+011824}\tablenotemark{1}\\
\hline\\[-0.25cm]		     						  					       	      	         
35 & 2.67 & 11:06:24.70\tablenotemark{7}  & $-$00:49:22.72 & 2.4152\tablenotemark{1}   & 19.816\tablenotemark{7} & 2002&APR&13&MMT      & 2.37 &    23 &     \object{2QZ J110624.6$-$004923}\tablenotemark{1}\\
36 &      & 11:06:35.14\tablenotemark{7}  & $-$00:50:03.42 & 2.4559\tablenotemark{1}   & 20.412\tablenotemark{7} & 2002&APR&14,15&MMT   & 2.38 & \phn3 &     \object{2QZ J110635.1$-$005004}\tablenotemark{1}\\ 
\hline\\[-0.25cm]		     						  						      	         
37 & 1.26 & 12:12:51.18\tablenotemark{7}  & $-$00:53:40.82 & 2.4727\tablenotemark{1}   & 20.007\tablenotemark{7} & 2003&MAR&07&MMT      & 2.95 & \phn7 &     \object{2QZ J121251.1$-$005342}\tablenotemark{1}\\
38 &      & 12:12:56.10\tablenotemark{7}  & $-$00:53:35.39 & 2.4586\tablenotemark{1}   & 20.427\tablenotemark{7} & 2003&MAR&07,08&MMT   & 2.89 & \phn4 &     \object{2QZ J121256.0$-$005336}\tablenotemark{1}\\
\hline\\[-0.25cm]		     						  						      	         
39 & 3.77 & 12:27:07.11\tablenotemark{1}  & $-$01:17:16.8  & 2.2143\tablenotemark{1}   & 19.937\tablenotemark{1} & 2003&MAR&23&MMT      & 2.37 & \phn8 &     \object{2QZ J122707.1$-$011718}\tablenotemark{1}\\
40 &      & 12:27:18.78\tablenotemark{7}  & $-$01:19:40.94 & 2.3368\tablenotemark{1}   & 19.452\tablenotemark{7} & 2003&MAR&23&MMT      & 2.34 &    10 &     \object{2QZ J122718.7$-$011942}\tablenotemark{1}\\
\hline\\[-0.25cm]		     						  					       	                    
41 & 2.91 & 13:09:41.95\tablenotemark{7}  & $-$02:23:57.58 & 2.6645\tablenotemark{1}   & 19.480\tablenotemark{7} & 2004&MAY&19&MMT      & 2.38 &    17 &     \object{2QZ J130941.9$-$022358}\tablenotemark{1}\\
42 &      & 13:09:42.16\tablenotemark{8}  & $-$02:26:52.34 & 2.5786\tablenotemark{8}   & 18.972\tablenotemark{8} & 2004&MAY&18&MMT      & 2.32 &    11 &     \object{2QZ J130942.1$-$022652}\tablenotemark{1}\\
   &      &                               &                &                           &                         &     &   &  &         &      &       &          SDSS J130942.15$-$022652.3\tablenotemark{7}\\
\hline\\[-0.25cm]		     						  					       	                    
43 & 1.27 & 13:27:58.84\tablenotemark{8}  & $-$02:30:25.46 & 2.3434\tablenotemark{8}   & 19.261\tablenotemark{8} & 2003&MAR&27&MMT      & 2.67 &    12 &     \object{2QZ J132758.8$-$023025}\tablenotemark{1}\\
   &      &                               &                &                           &                         &     &   &  &         &      &       &          SDSS J132758.83$-$023025.4\tablenotemark{7}\\
44 &      & 13:27:59.79\tablenotemark{7}  & $-$02:31:40.27 & 2.3604\tablenotemark{1}   & 19.792\tablenotemark{7} & 2003&MAR&27&MMT      & 2.69 & \phn8 &     \object{2QZ J132759.8$-$023140}\tablenotemark{1}\\ 
\hline\\[-0.25cm]		     						  						      	         
45 & 0.84 & 13:28:30.14\tablenotemark{7}  & $-$01:57:32.78 & 2.3706\tablenotemark{1}   & 19.620\tablenotemark{7} & 2003&JUN&24,26&MMT   & 2.36 & \phn6 &     \object{2QZ J132830.1$-$015732}\tablenotemark{1}\\
46 &      & 13:28:33.63\tablenotemark{7}  & $-$01:57:27.94 & 2.3605\tablenotemark{1}   & 20.053\tablenotemark{7} & 2003&JUN&23&MMT      & 2.33 &    10 &     \object{2QZ J132833.6$-$015727}\tablenotemark{1}\\
\hline\\[-0.25cm]		     						  					       	                    
47 & 0.92 & 13:41:14.96\tablenotemark{8}  & $+$01:09:06.72 & 2.4422\tablenotemark{8}   & 18.615\tablenotemark{8} & 2003&MAR&28&MMT      & 3.10 &    10 &       \object{2QZ J134114.9+010906}\tablenotemark{1}\\
   &      &                               &                &                           &                         &     &   &  &         &      &       &            SDSS J134114.95+010906.7\tablenotemark{7}\\
48 &      & 13:41:15.56\tablenotemark{7}  & $+$01:08:12.70 & 2.2073\tablenotemark{1}   & 18.904\tablenotemark{7} & 2003&JUN&22&MMT      & 2.34 & \phn9 &       \object{2QZ J134115.5+010812}\tablenotemark{1}\\
\hline\\[-0.25cm]		     						  						      
49 & 4.07 & 13:47:26.25\tablenotemark{7}  & $-$00:57:33.04 & 2.3758\tablenotemark{1}   & 19.654\tablenotemark{7} & 2002&APR&13,14&MMT   & 2.44 &    13 &     \object{2QZ J134726.2$-$005734}\tablenotemark{1}\\
50 &      & 13:47:42.07\tablenotemark{7}  & $-$00:58:30.51 & 2.5146\tablenotemark{1}   & 19.305\tablenotemark{7} & 2002&APR&14&MMT      & 2.41 &    21 &     \object{2QZ J134742.0$-$005831}\tablenotemark{1}\\
\hline\\[-0.25cm]		     						  					       	           
51 & 2.46 & 13:54:12.61\tablenotemark{8}  & $-$02:01:42.25 & 2.3167\tablenotemark{8}   & 19.120\tablenotemark{8} & 2003&JUN&26&MMT      & 2.36 &    12 &                        QUEST 116007\tablenotemark{9}\\
   &      &                               &                &                           &                         & 2004&MAY&18&MMT      &      &       &     \object{2QZ J135412.5$-$020143}\tablenotemark{1}\\
   &      &                               &                &                           &                         &     &   &  &         &      &       &          SDSS J135412.61$-$020142.2\tablenotemark{7}\\
52 &      & 13:54:20.18\tablenotemark{7}  & $-$02:03:18.62 & 2.2159\tablenotemark{1}   & 20.632\tablenotemark{7} & 2003&JUN&26&MMT      & 2.38 & \phn8 &     \object{2QZ J135420.1$-$020319}\tablenotemark{1}\\
\hline\\[-0.25cm]		     						  						      
53 & 3.20 & 14:11:13.39\tablenotemark{7}  & $+$00:44:51.73 & 2.2620\tablenotemark{1}   & 19.562\tablenotemark{7} & 2003&JUN&24,26&MMT   & 2.40 & \phn8 &       \object{2QZ J141113.3+004451}\tablenotemark{1}\\
   &      &                               &                &                           &                         & 2004&MAY&19&MMT      &      &       &                                                     \\
54 &      & 14:11:23.52\tablenotemark{8}  & $+$00:42:52.96 & 2.2669\tablenotemark{8}   & 18.176\tablenotemark{8} & 2003&JUN&23&MMT      & 2.33 &    11 &                             UM 645\tablenotemark{10}\\
   &      &                               &                &                           &                         &     &   &  &         &      &       &                       MRC 1408+009\tablenotemark{11}\\
   &      &                               &                &                           &                         &     &   &  &         &      &       &                       TXS 1408+009\tablenotemark{12}\\
   &      &                               &                &                           &                         &     &   &  &         &      &       &   \object{SDSS J141123.51+004252.9}\tablenotemark{7}\\
\hline\\[-0.25cm]		     						  						      
55 & 3.74 & 14:42:45.66\tablenotemark{8}  & $-$02:42:50.15 & 2.3258\tablenotemark{8}   & 19.567\tablenotemark{8} & 2003&JUN&23,24&MMT   & 2.40 & \phn4 &     \object{2QZ J144245.6$-$024251}\tablenotemark{1}\\
   &      &                               &                &                           &                         &     &   &  &         &      &       &          SDSS J144245.66$-$024250.1\tablenotemark{7}\\
56 &      & 14:42:45.75\tablenotemark{7}  & $-$02:39:05.76 & 2.5568\tablenotemark{1}   & 20.119\tablenotemark{7} & 2003&JUN&22&MMT      & 2.39 &    10 &     \object{2QZ J144245.7$-$023906}\tablenotemark{1}\\
\hline\\[-0.25cm]		     						  					       	      
62 & 1.02 & 21:42:22.23\tablenotemark{13} & $-$44:19:29.8  & 3.22\tablenotemark{13}    & 21.21\tablenotemark{13} & 2002&OCT&30&CLAY     & 2.58 & \phn8 &             \object{Q 2139$-$4433}\tablenotemark{13}\\
63 &      & 21:42:25.86\tablenotemark{13} & $-$44:20:18.1  & 3.230\tablenotemark{13}   & 18.80\tablenotemark{13} & 2002&OCT&30&CLAY     & 2.54 &    21 &                   LBQS 2139$-$4434\tablenotemark{14}\\
   &      &                               &                &                           &                         &     &   &  &         &      &       &             \object{Q 2139$-$4434}\tablenotemark{13}\\
\hline\\[-0.25cm]		     						  					       	      
64 & 0.55 & 21:43:04.09\tablenotemark{13} & $-$44:50:36.0  & 3.06\tablenotemark{13}    & 21.14\tablenotemark{13} & 2002&NOV&01&CLAY     & 2.54 & \phn6 &            \object{Q 2139$-$4504A}\tablenotemark{13}\\
65 &      & 21:43:07.01\tablenotemark{13} & $-$44:50:47.6  & 3.25\tablenotemark{13}    & 21.27\tablenotemark{13} & 2002&NOV&01&CLAY     & 2.54 &    11 &            \object{Q 2139$-$4504B}\tablenotemark{13}\\
\hline\\[-0.25cm]		     						  					       	           
66 & 3.64 & 21:52:25.85\tablenotemark{1}  & $-$28:30:58.2  & 2.7378\tablenotemark{1}   & 19.160\tablenotemark{1} & 2002&NOV&01&CLAY     & 2.49 &    15 &     \object{2QZ J215225.8$-$283058}\tablenotemark{1}\\
67 &      & 21:52:40.05\tablenotemark{1}  & $-$28:32:50.8  & 2.7140\tablenotemark{1}   & 19.624\tablenotemark{1} & 2002&NOV&01&CLAY     & 2.51 &    14 &     \object{2QZ J215240.0$-$283251}\tablenotemark{1}\\
\hline\\[-0.25cm]		     						  					       	                    
68 & 3.19 & 22:38:50.10\tablenotemark{1}  & $-$29:56:11.5  & 2.4635\tablenotemark{1}   & 19.406\tablenotemark{1} & 2002&NOV&02&CLAY     & 2.52 &    17 &     \object{2QZ J223850.1$-$295612}\tablenotemark{1}\\
69 &      & 22:38:50.93\tablenotemark{1}  & $-$29:53:00.6  & 2.3859\tablenotemark{1}   & 19.529\tablenotemark{1} & 2002&NOV&02&CLAY     & 2.52 &    16 &     \object{2QZ J223850.9$-$295301}\tablenotemark{2}\\
\hline\\[-0.25cm]		     						  					       	                    
70 & 1.81 & 22:40:18.16\tablenotemark{1}  & $-$29:40:29.3  & 2.5567\tablenotemark{1}   & 20.593\tablenotemark{1} & 2003&SEP&28&BAADE    & 2.51 &    10 &     \object{2QZ J224018.2$-$294029}\tablenotemark{1}\\
71 &      & 22:40:22.28\tablenotemark{1}  & $-$29:38:55.2  & 2.5430\tablenotemark{1}   & 20.462\tablenotemark{1} & 2003&SEP&28&BAADE    & 2.50 & \phn7 &     \object{2QZ J224022.3$-$293855}\tablenotemark{1}\\
\hline\\[-0.25cm]		     						  					       	      	         
74 & 2.21 & 22:41:52.14\tablenotemark{1}  & $-$28:41:04.9  & 2.2870\tablenotemark{1}   & 19.221\tablenotemark{1} & 2003&SEP&28&BAADE    & 2.49 &    15 &     \object{2QZ J224152.2$-$284105}\tablenotemark{1}\\
75 &      & 22:41:58.38\tablenotemark{1}  & $-$28:42:49.4  & 2.3678\tablenotemark{1}   & 19.580\tablenotemark{1} & 2003&SEP&28&BAADE    & 2.49 & \phn9 &     \object{2QZ J224158.4$-$284250}\tablenotemark{1}\\
\hline\\[-0.25cm]		     						  					       	      	         
76 & 3.78 & 23:03:01.65\tablenotemark{1}  & $-$29:00:27.2  & 2.5687\tablenotemark{1}   & 19.251\tablenotemark{1} & 2002&OCT&31&CLAY     & 2.55 & \phn9 &     \object{2QZ J230301.6$-$290027}\tablenotemark{1}\\
77 &      & 23:03:18.48\tablenotemark{1}  & $-$29:01:20.2  & 2.5863\tablenotemark{1}   & 19.566\tablenotemark{1} & 2002&OCT&31&CLAY     & 2.55 &    16 &     \object{2QZ J230318.4$-$290120}\tablenotemark{1}\\
\hline\\[-0.25cm]		     						  					       	      	         
78 & 3.04 & 23:19:31.71\tablenotemark{1}  & $-$30:24:36.5  & 2.3835\tablenotemark{1}   & 20.060\tablenotemark{1} & 2003&SEP&30&BAADE    & 2.53 &    11 &     \object{2QZ J231931.7$-$302436}\tablenotemark{1}\\
79 &      & 23:19:42.78\tablenotemark{1}  & $-$30:26:29.7  & 2.4728\tablenotemark{1}   & 19.376\tablenotemark{1} & 2003&SEP&29&BAADE    & 2.53 &    18 &     \object{2QZ J231942.7$-$302630}\tablenotemark{1}\\
\hline\\[-0.25cm]		     						  					       	      	         
80 & 2.35 & 23:26:03.52\tablenotemark{1}  & $-$29:37:40.3  & 2.3104\tablenotemark{1}   & 20.580\tablenotemark{1} & 2003&SEP&29&BAADE    & 2.51 & \phn8 &     \object{2QZ J232603.5$-$293740}\tablenotemark{1}\\
81 &      & 23:26:14.26\tablenotemark{1}  & $-$29:37:22.1  & 2.3874\tablenotemark{1}   & 19.132\tablenotemark{1} & 2003&OCT&01&BAADE    & 2.50 &    12 &     \object{2QZ J232614.2$-$293722}\tablenotemark{1}\\
\hline\\[-0.25cm]		     						  					       	      	         
82 & 0.87 & 23:28:00.70\tablenotemark{1}  & $-$27:16:55.5  & 2.3799\tablenotemark{1}   & 20.582\tablenotemark{1} & 2002&OCT&31&CLAY     & 2.56 & \phn9 &     \object{2QZ J232800.7$-$271655}\tablenotemark{1}\\
83 &      & 23:28:04.41\tablenotemark{1}  & $-$27:17:13.0  & 2.3640\tablenotemark{1}   & 20.435\tablenotemark{1} & 2002&OCT&31&CLAY     & 2.56 &    10 &     \object{2QZ J232804.4$-$271713}\tablenotemark{1}\\
\hline\\[-0.25cm]		     						  					       	           
84 & 1.84 & 23:31:00.48\tablenotemark{1}  & $-$28:39:46.2  & 2.4803\tablenotemark{1}   & 20.436\tablenotemark{1} & 2003&SEP&28&BAADE    & 2.50 &    10 &     \object{2QZ J233100.4$-$283946}\tablenotemark{1}\\
   &      &                               &                &                           &                         & 2003&SEP&29&BAADE    &      &       &                                                     \\
85 &      & 23:31:05.17\tablenotemark{1}  & $-$28:38:14.4  & 2.4777\tablenotemark{1}   & 20.607\tablenotemark{1} & 2003&SEP&29&BAADE    & 2.50 &    22 &     \object{2QZ J233105.1$-$283814}\tablenotemark{1}\\
\hline\\[-0.25cm]		     						  					       	           
88 & 2.48 & 23:56:43.70\tablenotemark{1}  & $-$29:23:29.4  & 2.5344\tablenotemark{1}   & 20.781\tablenotemark{1} & 2003&SEP&30&BAADE    & 2.51 &    10 &     \object{2QZ J235643.6$-$292329}\tablenotemark{1}\\
   &      &                               &                &                           &                         & 2003&OCT&01&BAADE    &      &       &                                                     \\
89 &      & 23:56:44.52\tablenotemark{1}  & $-$29:25:57.6  & 2.5392\tablenotemark{1}   & 20.831\tablenotemark{1} & 2003&SEP&30&BAADE    & 2.50 &    10 &     \object{2QZ J235644.4$-$292557}\tablenotemark{1}\\
\cutinhead{QSO Triplet}		     						  					       	           
57 & 2.44 & 16:25:48.00\tablenotemark{7}  & $+$26:44:32.64 & 2.467\tablenotemark{15}   & 18.542\tablenotemark{7} & 2002&APR&14&MMT      & 2.35 &    32 &           \object{KP 1623.7+26.8A}\tablenotemark{16}\\
58 & 2.91 & 16:25:48.80\tablenotemark{8}  & $+$26:46:58.77 & 2.5177\tablenotemark{8}   & 17.340\tablenotemark{8} & 2002&APR&13&MMT      & 2.42 &    69 &                    KP 1623.7+26.8B\tablenotemark{16}\\
   &      &                               &                &                           &                         &     &   &  &         &      &       &   \object{SDSS J162548.79+264658.7}\tablenotemark{7}\\
59 & 2.11 & 16:25:57.38\tablenotemark{8}  & $+$26:44:48.22 & 2.6016\tablenotemark{8}   & 19.095\tablenotemark{8} & 2002&APR&15&MMT      & 2.52 &    23 &                     KP 1623.9+26.8\tablenotemark{16}\\
   &      &                               &                &                           &                         &     &   &  &         &      &       &   \object{SDSS J162557.38+264448.2}\tablenotemark{7}\\
\cutinhead{Single QSOs}		     						  						           
03 & 3.51\tablenotemark{b} 		     						  						           
   & 00:17:10.31\tablenotemark{17} & $-$38:56:25.1  & 2.347\tablenotemark{18}   & 17.98\tablenotemark{17} & 2002&OCT&30&CLAY     & 2.55 &    14 &              \object{Q 0014$-$392}\tablenotemark{18}\\
\hline\\[-0.25cm]		     						  					       	           
05 & \nodata\tablenotemark{c} 	     						  					       	           
   & 00:43:58.80\tablenotemark{19} & $-$25:51:15.53 & 2.501\tablenotemark{20}   & 17.16\tablenotemark{21} & 2002&NOV&01&CLAY     & 2.52 &    30 &                               CT34\tablenotemark{22}\\
   & &                               &                &                           &                         &     &   &  &         &      &       &            PBP84 004131.1$-$260740\tablenotemark{20}\\
   & &                               &                &                           &                         &     &   &  &         &      &       &                   LBQS 0041$-$2607\tablenotemark{14}\\
   & &                               &                &                           &                         &     &   &  &         &      &       &                      UJ3682P$-$013\tablenotemark{23}\\
   & &                               &                &                           &                         &     &   &  &         &      &       & \object{2MASS J00435879$-$2551155}\tablenotemark{19}\\
\hline\\[-0.25cm]		     						  						           
25 & \nodata\tablenotemark{d} 	     						  						           
   & 07:29:28.56\tablenotemark{7}  & $+$25:24:51.84 & 2.303\tablenotemark{24}   & 17.958\tablenotemark{7} & 2002&DEC&28&MMT      & 2.44 &    30 &             FIRST J072928.4+252451\tablenotemark{25}\\
   & &                               &                &                           &                         &     &   &  &         &      &       &               87GB 072625.3+253009\tablenotemark{26}\\
   & &                               &                &                           &                         &     &   &  &         &      &       &                     GB6 J0729+2524\tablenotemark{27}\\
   & &                               &                &                           &                         &     &   &  &         &      &       &                NVSS J072928+252450\tablenotemark{28}\\
   & &                               &                &                           &                         &     &   &  &         &      &       &                    FBQS J0729+2524\tablenotemark{24}\\
   & &                               &                &                           &                         &     &   &  &         &      &       &   \object{2MASS J07292848+2524517}\tablenotemark{19}\\
\hline\\[-0.25cm]		     						  						      
27 & 4.58\tablenotemark{b} 		     						  						      
   & 08:45:57.69\tablenotemark{8}  & $+$44:45:46.00 & 2.2684\tablenotemark{8}   & 18.361\tablenotemark{8} & 2003&JAN&06&MMT      & 3.17 & \phn8 &   \object{SDSS J084557.68+444546.0}\tablenotemark{7}\\
\hline\\[-0.25cm]		     						  						      	   
30 & 4.42\tablenotemark{b} 		     						  						           
   & 10:05:10.58\tablenotemark{7}  & $-$00:43:23.40 & 2.4390\tablenotemark{1}   & 19.972\tablenotemark{7} & 2002&APR&15&MMT      & 2.36 &    11 &     \object{2QZ J100510.5$-$004324}\tablenotemark{1}\\
\hline\\[-0.25cm]		     						  						     
32 & \nodata\tablenotemark{e} 	     						  						           
   & 10:28:32.62\tablenotemark{1}  & $-$01:34:47.3  & 2.2937\tablenotemark{1}   & 20.610\tablenotemark{1} & 2002&DEC&28&MMT      & 2.45 & \phn5 &     \object{2QZ J102832.6$-$013448}\tablenotemark{1}\\
   & &                               &                &                           &                         & 2003&MAR&23&MMT      &      &       &                                                     \\
\hline\\[-0.25cm]		     												     
72 & \nodata\tablenotemark{e} 	     												           
   & 22:40:26.22\tablenotemark{8}  & $+$00:39:40.13 & 2.1098\tablenotemark{8,f} & 18.529\tablenotemark{8} & 2003&SEP&28&BAADE    & 3.09 &    12 &                        2237.9+0040\tablenotemark{29}\\
   & &                               &                &                           &                         &     &   &  &         &      &       &   \object{SDSS J224026.21+003940.1}\tablenotemark{7}
																     
\enddata															     

\tablecomments{The columns (left to right) are: object number, QSO pair separation, 
  right ascension, declination, redshift, apparent magnitude, observation date(s), 
  spectral resolution full width at half maximum, mean signal-to-noise per
  pixel in the ``pure'' \lya\ forest (redward of the QSO's \lyb/\osx\ emission line), 
  and previously published QSO designations.}

% The lines below were changed in consultation with Alison Compton at ApJ.  
% The original form follws (commented).
\tablenotetext{a}{Magnitude filters:
 \citet{2006ade162apjs38} -- sdss g$^{\prime}$ PSF; 
 \citet{2004cro349mnras1397} -- b$_j$; 
 \citet{1993gou88apjs53} -- V;
 \citet{1995maz31rmxaa119} -- B; 
 \citet{2005sch130aj367} -- sdss g$^{\prime}$ PSF; 
 \citet{1998sir495apj659} -- Gunn r;
 \citet{1995ver296aap665} -- B; 
 \citet{1999zam346aap731} -- Johnson B}
% \tablenotetext{a}{Magnitude filters -- 
%  (1)~b$_j$; 
%  (4)~Johnson B; 
%  (5)~B; 
%  (7)~sdss g$^{\prime}$ PSF; 
%  (8)~sdss g$^{\prime}$ PSF; 
% (13)~B; 
% (17)~Gunn r;
% (21)~V}
\tablenotetext{b}{The neighboring QSO was unobserved.}
\tablenotetext{c}{A published redshift for the neighboring QSO was
                      incorrect, but has been updated in
		      \citet{2003ver412aap399}.}
\tablenotetext{d}{\citet{2001ver374aap92} includes a non-existent neighboring
   QSO that is not included in \citet{2003ver412aap399}.}
\tablenotetext{e}{The published redshift for the neighboring QSO was
                      incorrect.}
\tablenotetext{f}{The \citet{2001ver374aap92} redshift for this QSO
is z=2.2, in accordance with the original sample criteria.}
\tablerefs{
 (1)~\citet{2004cro349mnras1397};
 (2)~\citet{1990boy243mnras1};
 (3)~\citet{1992zit256mnras349};
 (4)~\citet{1999zam346aap731};
 (5)~\citet{1995maz31rmxaa119};
 (6)~this paper; 
 (7)~\citet{2006ade162apjs38};
 (8)~\citet{2005sch130aj367};
 (9)~\citet{2004ren618apj184};
(10)~\citet{1981mac45apjs113};
(11)~\citet{1981lar194mnras693};
(12)~\citet{1996dou111aj1945};
(13)~\citet{1995ver296aap665};
(14)~\citet{1991mor102aj1627};
(15)~\citet{1989cro336apj550};
(16)~\citet{1978sra221apj468};
(17)~\citet{1998sir495apj659};
(18)~\citet{1993kor88apjs357};
(19)~\citet{2003cut2246ycat0};
(20)~\citet{1984poc210mnras373};
(21)~\citet{1993gou88apjs53};
(22)~\citet{1992maz24rmxaa147};
(23)~\citet{1987driphdt};
(24)~\citet{2000whi126apjs133};
(25)~\citet{1997whi475apj479};
(26)~\citet{1991bec75apjs1};
(27)~\citet{1996gre103apjs427};
(28)~\citet{1998con115aj1693};
(29)~\citet{1985cra90aj987}}
\end{deluxetable}
% \clearpage
% \end{landscape}

%\thispagestyle{empty}
\tabletypesize{\footnotesize}
\begin{deluxetable}{r@{}l l c c c c}
\tablecolumns{7}
\tablewidth{0pc}
\tablecaption{Resolution Measurement Lines.\label{tbl_lines}}
\tablehead{
\multicolumn{2}{c}{} & \colhead{$\lambda$} & \colhead{} & \colhead{} & \colhead{}\\
\multicolumn{2}{c}{Ion} & \colhead{(\AA)} & \colhead{MMT} & \colhead{Baade} & \colhead{Clay}}
\startdata

He&I  & 3819.66 &   --    &   --    & $\surd$ \\
He&I  & 3888.65 & $\surd$ &   --    &   --    \\
He&I  & 3964.73 &   --    &   --    & $\surd$ \\
He&I  & 4026.23 &   --    &   --    & $\surd$ \\
He&I  & 4120.92 &   --    &   --    & $\surd$ \\
He&I  & 4143.76 &   --    &   --    & $\surd$ \\
Ar&I  & 4259.36 & $\surd$ &   --    &   --    \\
Ar&I  & 4300.10 &   --    & $\surd$ &   --    \\
Ar&I  & 4510.73 & $\surd$ & $\surd$ &   --    \\
Ar&II & 4545.05 & $\surd$ & $\surd$ &   --    \\
Ar&II & 4579.35 & $\surd$ & $\surd$ &   --    \\
Ar&II & 4609.57 & $\surd$ &   --    &   --    \\
Ar&II & 4657.90 & $\surd$ & $\surd$ &   --    \\
He&I  & 4713.22 &   --    &   --    & $\surd$ \\
Ar&II & 4764.87 & $\surd$ &   --    &   --    \\
Ar&II & 4806.02 & $\surd$ & $\surd$ &   --    \\
Ar&II & 4847.81 & $\surd$ & $\surd$ &   --    

\enddata

\tablecomments{Check marks indicate which comparison lamp lines were used to 
measure spectral resolution for data taken at the three telescopes.}

\end{deluxetable}

\tabletypesize{\footnotesize}
\begin{deluxetable}{c l l c c c l}
\tablecolumns{7}
\tablewidth{0pc}
\tablecaption{QSO Catalog Corrections.\label{tbl_newz}}
\tablehead{
\colhead{\#} & \colhead{$\alpha$ (J2000)} & \colhead{$\delta$ (J2000)} &
\colhead{$z_{error}$} & \colhead{$z\pm\sigma_z$} & 
\colhead{Emission Lines Used\tablenotemark{a}} & \colhead{Alias}}
\startdata
12 & 01:50:47.6\tablenotemark{1}  & $-$42:37:40    & 2.30\tablenotemark{1}                         & \nodata\tablenotemark{b} & \nodata                                    &          \object{UJ3690P$-$114}\tablenotemark{1}\\
23 & 03:17:41.25\tablenotemark{2} & $-$53:11:58.7  & 2.33\tablenotemark{2}                         & 2.215$\pm$0.0044         & \ly, \nfv, \sitw, \onsitw, \ctw, \sifrofr  &                  \object{CT426}\tablenotemark{2}\\
26 & 08:45:58.56\tablenotemark{3} & $+$44:45:55.80 & 2.30\tablenotemark{4}                         & \nodata\tablenotemark{b} & \nodata                                    &                 \object{WEE 18}\tablenotemark{4}\\ 
31 & 10:28:27.15\tablenotemark{5} & $-$01:36:40.6  & 2.3100\tablenotemark{5}, 2.393\tablenotemark{6} & 1.609$\pm$0.0015         & \cfr, \cthr                              & \object{2QZJ102827.1$-$013641}\tablenotemark{5}\\
73 & 22:40:40.08\tablenotemark{3} & $+$00:40:25.09 & 2.2\tablenotemark{7}                          & 1.447$\pm$0.0053         & \cfr, \cthr                                &          \object{2238.1$+$0041}\tablenotemark{7}\\
86 & 23:43:21.5\tablenotemark{8}  & $+$01:22:43    & 2.35\tablenotemark{8}                         & 0.461$\pm$0.0011         & \mgtw, \nefva, \nefvb                      &              \object{BG CFH 25}\tablenotemark{8}\\
87 & 23:43:24.1\tablenotemark{8}  & $+$01:19:20    & 2.34\tablenotemark{8}                         & 1.714$\pm$0.0047         & \cfr, \cthr, \sithrfethr, \althr           &              \object{BG CFH 27}\tablenotemark{8}\\
\enddata

\tablecomments{The columns (left to right) are: object number, right ascension, declination, 
  previously published incorrect redshift(s), revised redshift, emission lines
  used for redshift determination, and previously published QSO designation.}
\tablenotetext{a}{$\lambda_{rest}$  \citep[\AA; ][]{2001van122aj549}:
  \ly~$-$~1216.25, \nfv~$-$~1239.85, \sitw\ $-$1265.22, \onsitw~$-$~1305.42,	                                          
  \ctw~$-$~1336.60, \sifrofr~$-$~1398.33, \cfr~$-$~1546.15, \althr~$-$~1856.76,                                          
  \sithrfethr~$-$~1892.64, \cthr~$-$~1907.30, \mgtw~$-$~2800.26, \nefva~$-$~3345.39,				    
  \nefvb~$-$~3425.66}								      				    
\tablenotetext{b}{This object is \emph{not} a QSO.}
\tablerefs{(1)~\citet{1987driphdt};
           (2)~\citet{1995maz31rmxaa119};
           (3)~\citet{2006ade162apjs38};
           (4)~\citet{1985wee57apjs523};
           (5)~\citet{2004cro349mnras1397};
           (6)~\citet{2006cop370mnras1804};
           (7)~\citet{1985cra90aj987};
           (8)~\citet{1983gas272apj411}}
\end{deluxetable}

\end{document}